\documentclass[11pt]{article}
\usepackage{graphicx} % Required for inserting images

\usepackage{amsmath}
\usepackage{amssymb}
\usepackage{amsfonts}
\usepackage{braket}
\usepackage{physics}
\usepackage{slashed}
\usepackage{multirow,bigdelim}
\usepackage{extarrows}
\usepackage{hyperref}
\usepackage{tikz}
\usetikzlibrary{positioning, intersections, calc, arrows.meta,math} %tikzのlibrary

\usepackage{appendix}
\usepackage{color}
\usepackage{comment}
\usepackage{authblk}
 \usepackage[centering]{geometry}
 \usepackage{cite}

\newcommand{\fix}[1]{\textcolor{red}{[#1]}}

\newcommand{\Lef}{Lefschetz }

\title{Computation of the index on orbifold from the Atiyah-Segal-Singer fixed point theorem}
\author[1]{Shoto Aoki \footnote{shotoaoki@g.ecc.u-tokyo.ac.jp}}
\affil[1]{\small \textit{
Graduate School of Arts and Sciences, University of Tokyo, Komaba, Meguro-ku, Tokyo 153-8902, Japan
}}

\author[2]{Maki Takeuchi\footnote{maki\_t@yamaguchi-u.ac.jp}}
\affil[2]{\small \textit{Graduate School of Sciences and Technology for Innovation, Yamaguchi University, Yamaguchi-shi, Yamaguchi 753-8512, Japan}}

\date{\today}

\begin{document}

\begin{flushright}
UT-Komaba/24-5 
\end{flushright}

{\let\newpage\relax\maketitle}
%\maketitle

\begin{abstract}
    We investigate the independent chiral zero modes on the orbifolds from the Atiyah-Segal-Singer fixed point theorem. The required information for this calculation includes the fixed points of the orbifold and the manner in which the spatial symmetries act on these points, unlike previous studies that necessitated the calculation of zero modes. Since the fixed point theorem can be applied to any fermionic theory on any orbifold, it allows us to determine the index even on orbifolds where the calculation of zero modes is challenging or in the presence of non-trivial gauge configurations.
    We compute the indices on the $T^{2}/ \mathbb{Z}_N\,(N=2,3,4,6)$ and $T^{4}/ \mathbb{Z}_N\,(N=2,3,5)$ as examples. Furthermore, we also attempt to compute the indices on a Coxeter orbifold related to the $D_4$ lattice.
\end{abstract}

\newpage
\tableofcontents
\newpage

\section{Introduction}
%余剰次元模型
% One of the mysteries of the Standard Model is why there are three identical copies of quarks and leptons, each with the same charge and spin quantum numbers. 
Higher-dimensional theories are strong candidates for beyond the Standard Model and can potentially explain the three generations of quarks and leptons \cite{Sakamoto:2020pev,Abe:2008sx,Abe:2015yva,Libanov:2000uf,Frere:2000dc,PhysRevD.65.044004,PhysRevD.73.085007,Gogberashvili:2007gg,Guo:2008ia,PhysRevLett.108.181807}, the origin of the flavor structure, such as mass hierarchy \cite{Cremades:2004wa,Abe:2008sx,Arkani-Hamed:1999ylh,Dvali:2000ha,Gherghetta:2000qt,Kaplan:2000av,Huber:2000ie,Kaplan:2001ga,Fujimoto:2012wv,PhysRevD.97.115039,PhysRevD.90.105006} and CP violation\cite{PhysRevD.88.115007,PhysRevD.90.105006,Kobayashi:2016qag,Buchmuller:2017vho,PhysRevD.97.075019}. In particular, toroidal orbifold compactifications \cite{Dixon:1985jw,Dixon:1986jc} have been extensively studied as a means to obtain the chiral spectrum.

%普通の指数
For constructing a realistic model, it is crucial to be able to obtain chiral fermions in the model, and the Atiyah-Singer (AS) index theorem \cite{Atiyah:1963zz} is useful as a tool.
% The Atiyah-Singer (AS) index theorem is the powerful tool to obtain the number of chiral fermions. 
The index of a Dirac operator $\slashed{D}$
\begin{align}
    {\rm{Ind}}(i\slashed{D})\equiv n_{+}-n_{-}
    \label{index1}
\end{align}
is a topological invariant. Here, $n_{\pm}$ are the numbers of right-handed and left-handed zero modes for the Dirac operator. 
%In general, the AS index theorem is used for smooth manifolds. 
The AS index theorem for a $2D$ compact manifold ${\mathcal{M}}^2$ with magnetic flux is known as \cite{Witten:1984dg,Green:1987mn}
\begin{align}
     n_{+}-n_{-}=\frac{1}{2\pi}\int_{{\mathcal{M}}^2} F
     \label{index2}
\end{align}
where $F$ is a $2$-form field strength of the flux. If we take ${\mathcal{M}}^2$ to be a $2D$ torus $T^2$ or a $2D$ sphere $S^2$, the index $n_{+}-n_{-}$ is equal to magnetic flux quantization number $M$. 
% In these models, we can obtain three generations only in the case $M=3$.

%Orbifold上の指数
On the other hand, extra-dimensional models with singularities are often valuable. In particular, orbifold models, which possess singularities, exhibit a rich structure, making them powerful tools for reproducing phenomena such as the three generations \cite{Sakamoto:2020pev,Abe:2008sx,Abe:2015yva,Abe:2013bca,Abe:2014noa,Kobayashi:2017dyu,Kobayashi:2022tti,Imai:2022bke,Kikuchi:2022lfv,Kikuchi:2022psj,Kikuchi:2023awm} and flavor structures \cite{PhysRevD.88.115007,Fujimoto:2016zjs,Kobayashi:2016qag,Buchmuller:2017vho,PhysRevD.97.075019,Kikuchi:2021yog,Hoshiya:2022qvr} in phenomenological models.
The number of chiral zero modes on the magnetized orbifolds $T^2/{\mathbb{Z}}_N\,(N=2,3,4,6)$ has been investigated in \cite{Abe:2013bca,Abe:2014noa,Kobayashi:2017dyu}. However, a complete list of the chiral zero mode numbers on these orbifolds has yet to be compiled.
The complete list of them %the number of chiral zero modes on the $T^2/{\mathbb{Z}}_N\,(N=2,3,4,6)$ with magnetic flux 
has been found in \cite{Sakamoto:2020pev}, and later derived as the AS index theorem on the orbifold in \cite{Sakamoto:2020vdy,Kobayashi:2022tti,Imai:2022bke}.
% The index theorem on $T^2/{\mathbb{Z}}_N\,(N=2,3,4,6)$ with magnetic flux is studied in \cite{Sakamoto:2020pev,Sakamoto:2020vdy,Kobayashi:2022tti,Imai:2022bke}, 
The index on $T^2/{\mathbb{Z}}_N\,(N=2,3,4,6)$ with magnetic flux  are given as \cite{Kobayashi:2022tti}
\begin{align}
 n_{+}-n_{-}=\frac{M}{N}+\sum_{I}\frac{\xi_{I}^{F}}{N},
 \label{indT2ZN}
\end{align}
where $M$ is flux quantization number and $\xi_{I}^{F}$ is the localized flux at the fixed points of $T^2/{\mathbb{Z}}_N$ orbifold. $I$ is the label of the fixed points of $T^2/{\mathbb{Z}}_N$ orbifold.
This formula reveals that, in comparison to the index of smooth manifolds, additional contributions from fixed points emerge.
In the context of superstring theory, it is intriguing to derive the indices in $T^4/{\mathbb{Z}}_N$ and $T^6/{\mathbb{Z}}_N$ orbifold models, which are considered useful. Although the expressions have been demonstrated in certain cases~\cite{Kikuchi:2022lfv,Kikuchi:2022psj,Kikuchi:2023awm}, a general proof remains elusive. %it has not been shown in general cases.

%Atiyah-Segal-Singer fixed point
In order to compute orbifold indices for general cases, we focus on the Atiyah-Singer index with a group action \cite{atiyah1967lefschetz,Atiyah1968TheIndex, berline1985computation,bismut1985infinitesimal,bismut1986localization}.
%An orbifold is constructed from a manifold with group action. 
We assume that the action commutes with the Dirac operator and the chiral operator. Then, the eigenspace of the Dirac operator is more finely classified by the eigenvalue (or the representation) of the group. In particular, the Dirac zero modes are characterized by the chirality and the eigenvalue of the group action. We define the \Lef number by
%Atiyah and Segal \cite{Atiyah1968TheIndex} defined the Lefschetz number as 
the difference between the trace of the group action on the chiral positive and negative modes. Atiyah and Bott \cite{atiyah1967lefschetz} have expressed the \Lef number in terms of de Rahm cohomology when all the fixed points are isolated and Atiyah, Segal, and Singer \cite{Atiyah1968TheIndex} have generalized this expression for general cases. 
We will show that the indices on an orbifold are the Fourier transformation of the Lefschetz numbers.

%論文のアピール
In this paper, we employ the Atiyah-Bott and Atiyah-Segal-Singer fixed point theorem, which is applicable to models with general singularities. We demonstrate that the contributions from fixed points can be expressed by the Lefschetz number. This allows a relatively straightforward derivation of indices in models with singularities and provides insight into the correspondence between the Lefschetz number and physical quantities.

%論文の構成
This paper is organized as follows. In Section \ref{Section2}, we review the Atiyah-Bott and Atiyah-Segal-Singer fixed point theorem and the \Lef number. In Section \ref{sec:Setup}, we briefly explain the setup of the magnetized orbifold model. In Section \ref{T2ZN}, we calculate the index on the $T^2/{\mathbb{Z}}_N\,(N=2,3,4,6)$ orbifolds with magnetic flux by using Atiyah-Bott fixed point theorem. 
After confirming that these results coincide with those of papers \cite{Sakamoto:2020pev,Kobayashi:2022tti,Imai:2022bke}, we proceed to elucidate the relationship between the Lefschetz number and physical quantities at the fixed points of the $T^2/{\mathbb{Z}}_N$ orbifold.
In Section \ref{T4ZN}, we evaluate the index on the $T^4/{\mathbb{Z}}_N$ orbifolds with magnetic flux by using the Atiyah-Bott fixed point theorem. In Section \ref{sec:Coxeter}, we consider the $D_4$ root lattice and orbifold constructed by the Coxeter element, which is not written by module transformation. Section \ref{Conclusion} is devoted to the discussion and conclusion. 

\section{Dirac operator with a group action}
\label{Section2}
%この章ではorbifold 指数とLefschets数の関係について議論する。orbifold上でDirac演算子の完全系を構成せずに、固定点の情報から抽出できることを示す。

We discuss the relationship between the index on an orbifold and \Lef number. %The index depends on the topology of the orbifold, while the Lefschets number is defined by just fixed points. 
We show that it is possible to extract information from fixed points without constructing a complete system of a Dirac operator on an orbifold.

%In this section, we start from a Riemannian manifold with a group action. 

%Let $g=g_{\mu \nu} dx^{\mu} dx^{\nu}$ be a metric on $X$. The action of the Lie group does not change the metric $g$. We can identify 

%Orbifold is obtained from a Riemannian manifold with a group action. 

\subsection{Index on Orbifold}
%群作用

%Orbifold と Manifoldの関係
An orbifold is constructed from a Riemannian manifold. Let $M$ be a $2n$-dimensional Riemannian manifold $M$, and $g=g_{\mu \nu} dx^{\mu} dx^{\nu}$ be a metric on $M$. We denote an inverse matrix of $g_{\mu\nu}$ by $g^{\mu \nu}$. Here, we assume that a finite Abelian group $H$ of order $N$ acts on $M$ and the action preserves the metric $g$. Such a manifold $M$ is called $H$-space. The action of $H$ on $M$ is denoted by $ h \cdot x $ for $h \in H$ and $x \in M$. %The action of the Lie group does not change the metric $g$. 

For each point $x \in M$, the subset
\begin{align}
    H \cdot x = \Set{ h \cdot x | h\in H} 
\end{align}
is called the orbit of $x$. If two orbits $H\cdot x$ and $H \cdot y$ cross each other, $H\cdot x$ is equal to $H \cdot y$. Then, we can define an equivalence relation $x \sim y$ if and only if $H \cdot x= H \cdot y$. %Namely, there exists $h \in H$ such that $y=h\cdot x$. 
The orbifold is the set of all orbits, denoted by $M/H$ \footnote{In mathematics, $H\backslash M$ is more appropriate rather than $M/H$.}.

Let $\pi:E\to M$ be a complex vector bundle over an $H$-space $M$. If $E$ is also an $H$-space and $\pi $ satisfies
\begin{align}
    \pi ( \rho(h) \cdot e)= h \cdot \pi (e),
\end{align}
Then, $E$ is called an $H$-vector bundle. Here, $\rho$ is a representation of $H$ and $\rho(h) :E \to E$ is an isomorphism. For $x \in X$, the restriction of $\rho(h) :E \to E$ is a linear isomorphism $\rho(h,x):E_x \to E_{h \cdot x} $. For $h_1,~h_2 \in H$, the composition of $\rho(h_1)$ and $\rho(h_2)$ is 
\begin{align}
    \rho(h_1) \circ \rho(h_2)=\rho(h_1 h_2)
\end{align}
and restricts to 
\begin{align}
    \rho(h_1, h_2 x) \rho(h_2,x)=\rho(h_1 h_2,x)
\end{align}
on $E_x$. For the identity element $1_H \in H$, $\rho(1_H)$ is the identity map on $M$.

We fix $h \in H$. We take a local trivial neighborhood $\qty{U_\alpha}$ on $M$ and assume that each $U_\alpha$ does not change under the action of $h$. Namely, $h(U_\alpha) \subset U_\alpha$. Let $\qty{t_{\alpha \beta}}$ be a set of transition functions. We denote a restriction of $\rho(h)$ to $\pi^{-1}(U_\alpha) \simeq U_\alpha \times \mathbb{C}^n$ by $\rho_\alpha (h)$. When $x \in U_\alpha \cap U_\beta $, $\rho_\alpha (h)$ and $\rho_\beta (h)$ must satisfy
\begin{align}
    \rho_\alpha (h,x )t_{\alpha \beta}(x)= t_{\alpha \beta}(h x) \rho_\beta (h,x) .
\end{align}

Let $\mathbf{S}$ be a spinor bundle over $M$, and let $V$ be a complex vector bundle with gauge group $G$. $V$ represents a degree of freedom of color. We assume that $\mathbf{S} \otimes V$ is an $H$-bundle and $\rho$ denotes a representation of $H$ on $\mathbf{S} \otimes V$\footnote{It is a nontrivial problem whether or not the symmetry on the base space lifts to the spinor bundle.}. A Dirac fermion $\psi$ is defined as a section $M \to  \mathbf{S} \otimes V$, and the action of $h \in H$ is given by
\begin{align}\label{eq:action for section}
    (h\psi)(x) = \rho(h,h^{-1} x) \psi( h^{-1} x).
\end{align}
\begin{comment}
$\rho $ is a representation of $h$ so $\rho(e, x)$ is equal to the identity map on $X$. For $h_1,~h_2 \in H$, the composition law is
\begin{align}
    \rho( h_1, h_2 \cdot x) \rho(h_2 ,x)= \rho(h_1 h_2,x).
\end{align}
If $x$ is a fixed point, $\rho$ is an ordinary representation of a finite-dimensional vector space. 
\end{comment}

On the tangent space at $x \in M$, we choose a orthonormal basis $\qty{e^\mu_a}$ of $g_{\mu\nu }(x)$:
\begin{align}
    g_{\mu \nu}(x) e^\mu_a e^\nu_b= \delta_{ab}.
\end{align}
When $n$ independent vector fields $e^\mu_1,\cdots,e^\mu_n $ are smooth around $x \in M$, the set of the vector fields is called a vielbein. The subscripts $a$ and $b$ run from $1$ to $n$ and represent the local Lorentz symmetry $SO(n)$. We denote $e^a_\mu$ as the inverse matrix of $e^\mu_a $. 

Let $\gamma^a$ be a gamma matrix such that $\qty{\gamma^a, \gamma^b }=2 g^{a b}$. We also define $\Gamma^\mu(x)= \gamma^a e_a^\mu(x)$, which is a gamma matrix along $e_a^\mu$ and satisfies
\begin{align}
    \qty{\Gamma^\mu (x) , \Gamma^\nu(x)}=2 g^{\mu \nu}(x).
\end{align}
The Dirac operator for the Dirac spinor $\psi$ is written as
\begin{align}
    i\slashed{D} = i\Gamma^\mu \qty(\pdv{}{x^\mu} + \frac{1}{4} \sum_{b,c} \omega_{bc,\mu} \gamma^b \gamma^c -i A_\mu ) ,
\end{align}
where the second term is a spin connection and the third term is a gauge connection of $G$. Since $M$ is a compact manifold, the Hilbert space $\mathcal{H}$ can be decomposed as
\begin{align}
 \mathcal{H}=\bigoplus_{\lambda \in \text{Spec}(i\slashed{D})} \mathcal{H}_\lambda,
\end{align}
where $\text{Spec}(i\slashed{D})$ is a set of all eigenvalues of $i\slashed{D}$ and 
\begin{align}
    \mathcal{H}_\lambda= \Set{ \psi \in \mathcal{H} |i \slashed{D} \psi= \lambda \psi }
\end{align}
is a finite-dimensional vector space.

The Dirac operator anti-commutes with a chiral operator
\begin{align}
    \bar{\gamma}=(-i\gamma^1 \gamma^2) \cdots (-i\gamma^{2n-1} \gamma^{2n}). 
\end{align}
It maps $\mathcal{H}_\lambda$ to $\mathcal{H}_{-\lambda}$. If the Dirac operator $\slashed{D}$ has zero modes, $\bar{\gamma}$ is a closed map on $\mathcal{H}_0$. Then, $\mathcal{H}_0$ is divided into two parts
\begin{align}
    \mathcal{H}_0= \mathcal{H}^+_0 \oplus \mathcal{H}^-_0, 
\end{align}
where the upper subscripts denote the eigenvalue of $\bar{\gamma}$.

%We assume that the action of $H$ commutes with $i\slashed{D} $ and $\bar{\gamma}$.

Assuming that the action of $ H$ commutes with both the Dirac operator $\slashed{D}$ on $M$ and the chiral operator $\bar{\gamma}$. %Then, each $\mathcal{H}_\lambda$ is a representation space of $H$ induced by \eqref{eq:action for section}. 
The action \eqref{eq:action for section} induces a representation $\rho_\lambda$ on $\mathcal{H}_\lambda$ and $\rho_0^\pm $ on $\mathcal{H}_0^\pm$. 
%We write the representation as $\rho_\lambda$ for $\mathcal{H}_\lambda~(\lambda\neq 0)$ and $\rho_0^\pm $ for $\mathcal{H}_0^\pm$. 
To emphasize $\mathcal{H}_\lambda$ and $\mathcal{H}_0^\pm$ are equipped with the representations $\rho_\lambda$ and $\rho_0^\pm $, we write these representation spaces as $(\mathcal{H}_\lambda, \rho_\lambda)$ and $(\mathcal{H}_0^\pm, \rho_0^\pm)$. 
Through block diagonalization, they can be decomposed into irreducible representation spaces as
\begin{align}
    (\mathcal{H}_\lambda ,\rho_\lambda)& \simeq \bigoplus_{i } ( V_i  ,\rho_i)^{\oplus C_{\lambda,i}}~(\lambda \neq 0), \\
    (\mathcal{H}_0^{\pm} ,\rho_0^\pm)& \simeq \bigoplus_{i } ( V_i  ,\rho_i)^{\oplus C_{0,i}^{\pm}},
\end{align}
where each $(V_i,\rho_i)$ is an irreducible representation space, and $i$ is a label. Since $H$ is Abelian, the dimension of $V_i$ is one, and $\rho_i(h) \in \mathbb{Z}_N \subset \mathbb{C}$ for all $h\in H$. $C_{\lambda,i}$ and $C_{0,i}^\pm$ denote the number of $(V_i, \rho_i)$ contained in $(\mathcal{H}_\lambda ,\rho_\lambda)$ and $(\mathcal{H}_0^\pm ,\rho_0^\pm)$, respectively. These numbers are non-negative integers, but most of them are zero except for finitely many $i$. Since the action of $H$ commutes with the chiral operator, $C_{-\lambda,i}$ is equal to $C_{\lambda,i}$ Then, the orbifold index for $\rho_i$ is defined by
\begin{align}
    \text{ind}_{M/H} ( \rho_i)= (C_{0,i}^+ - C_{0,i}^-).
\end{align}
%\fix{この定義は議論の余地があります。orbifoldがmanifoldになる特別な場合と整合した定義にするべきかもしれません}\\
%$\mathcal{H}_{\lambda,i}$ is a zero-dimensional vector space except for finitely many $i$. 

%orbifoldの指数

\subsection{Lefschetz number}
%Lefschetz number

%In mathematics, the index for an equivalent $H$-vector bundle has been studied for a long time. 

%\Lef number has been studied for a long time. 

%\fix{equaivalent H bundleの数学側の説明。Lefschetz数につながるようにかく}

In this section, we express the orbifold index $\text{ind}_{M/H} ( \rho_i)$ in terms of a characteristic class. We define a \Lef number of $h \in H$ as the trace of $\bar{\gamma} h$:
\begin{align}\label{eq:DefOfLef}
    L(h):=\tr( \bar{\gamma} h)= \sum_{i}(C_{0,i}^+ - C_{0,i}^-) \tr_{V_i} (\rho_i(h))= \sum_{i}(C_{0,i}^+ - C_{0,i}^-) \rho_i(h).
\end{align}
When $h=1_H$, this equation is equal to the Dirac index. Let $h_1$ be a generator of $H$. Then, $H= \set{h_1^k| k=0,1,\cdots ,N-1}$ and 
\begin{align}
    \sum_{k=0}^{N-1} \rho_i(h_1^k )^\ast \rho_j(h_1^k)= N \delta_{ij}
\end{align} 
holds by the orthogonality of irreducible representations. Here, $\rho_i(h_1^k )^\ast$ is the complex conjugate of $\rho_i(h_1^k )$. Since $\rho_i(h_1^k )$ is an element of $\mathbb{Z}_N \subset \mathbb{C}$, this equation is equivalent to the discrete Fourier transformation. Thus, the orbifold index for $\rho_i$ is given by
\begin{align}
    \text{ind}_{M/H} ( \rho_i)=  \frac{1}{N} \sum_{k=0}^{N-1} \rho_i(h_1^k )^\ast L(h_1^k) \label{eq:FT of LEf number} % \int_H \overline{ \chi_{\rho_i}(h) } L(h) ~d\mu (h).%=\text{dim}(V_i) \int_H \overline{ \chi_{\rho_i}(h) } \tr( \bar{\gamma} h) ~d\mu (h)
\end{align}

\begin{comment}

On the right hand side, $\tr_{V_i} (\rho_i(h))$ is called a character of a irreducible representation $\rho_i$. Let $\chi_{\rho}$ be a character of a representation $\rho$, then the characters of an irreducible representation are orthogonal to each other: 
\begin{align}
    \int_H \overline{ \chi_{\rho_j}(h) }\chi_{\rho_i} (h) ~d\mu(h)= \delta_{ij},
\end{align}
where $\rho_i$ and $\rho_j$ are irreducible representations and $\overline{ \ast }$ means a simple complex conjugate. $d\mu $ denotes a Haar measure normalized by $1$. This theorem is called Schur orthogonality relations. Then, the orbifold index is given by
\begin{align}
    \text{ind}_{M/H} ( \rho_i)= \int_H \overline{ \chi_{\rho_i}(h) } L(h) ~d\mu (h).%=\text{dim}(V_i) \int_H \overline{ \chi_{\rho_i}(h) } \tr( \bar{\gamma} h) ~d\mu (h)
\end{align}
\end{comment}

According to the Atiyah-Segal-Singer fixed point theorem, $L(h)$ is expressed by an integral of a characteristic class over the set of fixed points. Let $M^h$ be a subset of $M$ defined by
\begin{align}
    M^h=\Set{x \in M | h\cdot x=x }.
\end{align}
We assume that $\dim M^h=2n^\prime$ are even. Since $h$ does not change the base space $M^h$, $h$ leads to a linear isomorphism on each fiber of $TM |_{M^h}$. The linear map preserves the metric, making it an element of $SO(n)$. The tangent bundle of $TM^h$ is a subbundle of $TM |_{M^h}$ and equivalent to the eigenspace of $h=+1$. Then, $TM |_{M^h}$ is divided into two parts
\begin{align}
    TM |_{M^h}=TM^h \oplus \mathcal{N}.
\end{align}
$\mathcal{N}$ is equivalent to the normal bundle of $M^h$, and the rank is $2n^{\prime \prime}=2n-2n^\prime$. $\mathcal{N}$ can also be interpreted as a tubular neighborhood of $M^h$ within $M$. Since $\mathcal{N}$ is a fiber bundle over $M^h$, we take a coordinate $x=(x^\prime, x^{\prime \prime})$, where $x^\prime$ is a local coordinate on $M^h$ and $x^{\prime \prime}$ denotes the fiber direction.

Let $R$ be a Riemannian curvature tensor on $M$. Then, the restriction to $M^h$ is given by
\begin{align}
    R|_{M^h}= \mqty( R^{\prime} & \ast \\ \ast & R^{\prime \prime}),
\end{align}
where $R^\prime$ is a Riemannian curvature tensor on $M^h$ and $R^{\prime \prime}$ is a curvature tensor of $\mathcal{N}\to M^h$.

The spinor bundle $\mathbf{S}$ is also divided as
\begin{align}
    \mathbf{S}|_{M^h}= \mathbf{S}^{\prime}  \otimes \mathbf{S}^{\prime \prime}
\end{align}
on $M^h$. $\mathbf{S}^{\prime }$ is the spinor bundle of $M^h$ and $\mathbf{S}^{\prime \prime}$ is the spinor bundle constructed by the spin structure of $\mathcal{N}$. Let $\gamma^{\prime i}~(i=1,\cdots,2n^\prime)$ and $\gamma^{\prime\prime a}~(a=1,\cdots,2n^{\prime\prime})$ be a gamma matrix for $\mathbf{S}^\prime$ and $\mathbf{S}^{\prime\prime}$. The gamma matrix for $\mathbf{S}$ is written as
\begin{align}
    \gamma^I= \left\{ \begin{array}{ll}
        \gamma^{\prime i} \otimes \bar{\gamma}^{\prime\prime} & (I=i) \\
        1 \otimes \gamma^{\prime \prime a} & (I=2n^\prime +a)
    \end{array}, \right.
\end{align}
where $\bar{\gamma}^{\prime \prime}$ is a chirality operator for $\mathbf{S}^{\prime\prime}$. The action of $h$ does not change $\mathbf{S}^\prime$, so $\rho(h)$ does not include $\gamma^{\prime i}$. Now we get the curvature tensor for $\mathbf{S}|_{M^h}$ as
\begin{align}
R|_{M^h}= &R_{IJ kl} \frac{ \gamma^I \gamma^J }{4} \frac{dx^{\prime k} dx^{\prime l}}{2} \nonumber \\
=&  R_{ij kl }^\prime \frac{ \gamma^{\prime i} \gamma^{\prime j} \otimes 1  }{4}\frac{dx^{\prime k} dx^{\prime l} }{2} +   R_{ab kl}^{\prime\prime} \frac{ 1 \otimes \gamma^{\prime \prime a} \gamma^{\prime \prime b } }{4} \frac{dx^{\prime k} dx^{\prime l} }{2} +( \text{mixing terms}) . 
\end{align}
Here, we can ignore the mixing terms between $TM^h $ and $\mathcal{N}$.
We let $\slashed{R}^{\prime \prime}=R_{ab kl}^{\prime\prime} \frac{ \gamma^{\prime \prime a} \gamma^{\prime \prime b } }{4} \frac{dx^{\prime k} dx^{\prime l} }{2}$. Then, $L(h)$ is obtained as
\begin{align}
    L(h)= \int_{M^h} \hat{A} (M^h) \frac{ \text{tr}_{\mathbf{S}^{\prime \prime} \otimes V} \qty( \rho(h) \bar{\gamma}^{\prime\prime} e^{\frac{i}{2\pi} \slashed{R}^{\prime\prime} }  e^{\frac{1}{2\pi} F}) }{  \det_\mathcal{N}( 1- e^{\frac{i}{2\pi} R^{\prime\prime}} h)} \label{eq:AtiyahSegalSinger},
\end{align}
where $\hat{A} (M^h)$ is an $A$-roof class and $F$ is a curvature tensor for $V$. This nontrivial relation is well-known as the Atiyah-Segal-Singer index theorem \cite{Atiyah1968TheIndex, berline1985computation,bismut1985infinitesimal,bismut1986localization}. The first proof has been given by $K$-theory in \cite{Atiyah1968TheIndex}. Later, some mathematicians have proven this theorem by the heat kernel method in \cite{berline1985computation,bismut1985infinitesimal,bismut1986localization}
%. This method is so sophisticated, but too abstract and not for physicists. 

%In appendix \ref{app:InfLefNum} and \ref{app:Localization}, we introduce a heat kernel method when $h$ is associated with an infinitesimal transformation \cite{bismut1985infinitesimal} and localization method to the fixed points \cite{bismut1986localization}.

If $M^h$ is a set of isolated points, the expression of $L(h)$ becomes simpler. $\mathbf{S}$ is equivalent to $\mathbf{S}^{\prime \prime}$ at the fixed point and $\bar{\gamma}=\bar{\gamma}^{\prime \prime}$. Since the dimension of $M^h$ is zero, $\hat{A} (M^h), ~e^{\frac{i}{2\pi} \slashed{R}^{\prime\prime} } ,~ e^{\frac{1}{2\pi} F} $ and $e^{\frac{i}{2\pi} R^{\prime\prime}}$ are all equal to one. Now we get the Atiyah-Bott fixed point theorem \cite{atiyah1967lefschetz}:
\begin{align}
    L(h)= \sum_{x \in M^h} \frac{ \text{tr}_{\mathbf{S} \otimes V} \qty( \rho(h) \bar{\gamma}  )}   {  \det_\mathcal{N}( 1-  h)}. \label{eq:AtiyahBott}
\end{align}
Since there is no eigenstate with $h=1$ in $\mathcal{N}$, $\det_\mathcal{N}(1-h)$ is not zero. In the rest of this section, we prove this theorem.

The trace in \eqref{eq:DefOfLef} is defined on a function space. We take a basis $\Set{ \ket{x} | x \in M}$ satisfying
\begin{align}
    \bra{x} \ket{y}=\frac{1}{\sqrt{g}} \delta^{n}(x-y).
\end{align}
for $x,~y \in M$. Here, we assume that $x$ and $y$ are in the same coordinate neighborhood and the difference between $x$ and $y$ is defined on the same coordinate. Then, the \Lef number is given by
\begin{align}
    L(h)=& \tr (\bar{\gamma} h)= \lim_{t\to 0} \tr (\bar{\gamma} h e^{t \slashed{D}^2}) \nonumber \\
    =&   \lim_{t\to 0} \int_M dx^{n} \sqrt{g} ~\text{tr}_{\mathbf{S} \otimes V} \qty(  \bar{\gamma}   \bra{x} h e^{t \slashed{D}^2} \ket{x}) \nonumber \\
    =&  \lim_{t\to 0} \int_M dx^n \sqrt{g} ~\text{tr}_{\mathbf{S} \otimes V} \qty(  \bar{\gamma}  \rho(h,hx^{-1}) \bra{h^{-1}x}  e^{t \slashed{D}^2} \ket{x}).
\end{align}
When $t$ is small enough, $\bra{h^{-1}x}  e^{t \slashed{D}^2} \ket{x}$ looks like $\delta^n(h^{-1}x-x)$. We can ignore a contribution from $hx\neq x$ and consider it around the fixed points. Let $U_p$ be a small neighborhood around the fixed point $p\in M^h$, the integral over $M$ becomes to the summation of the integral over $U_p$ and we get
\begin{align}
    L(h)= \lim_{t\to 0} \sum_{p\in M^h} \int_{U_p} dx^n \sqrt{g} ~\text{tr}_{\mathbf{S} \otimes V} \qty(  \bar{\gamma}  \rho(h,hx^{-1}) \bra{h^{-1}x}  e^{t \slashed{D}^2} \ket{x}).
\end{align}
We take a geodesic coordinate around the fixed point $p\in M^h$ on $U_p$ as $x$. $x=0$ means $p$. Then, the metric $g$ and volume form $\sqrt{g}$ are expanded as
\begin{align}
    g_{\mu \nu}(x)& = \delta_{\mu \nu }- \frac{1}{3} R_{\mu \rho \nu \sigma} (p) x^\rho x^\sigma +\order{x^3} , \\
    \sqrt{g} &= 1- \frac{1}{3}\text{Ric}_{\mu \nu} (p) x^\mu x^\nu +\order{x^3}.
\end{align}
The heat kernel $ \bra{h^{-1}x}  e^{t \slashed{D}^2} \ket{x}$ is also expanded
\begin{align}
     \bra{h^{-1}x}  e^{t \slashed{D}^2} \ket{x}=& \frac{1}{ \qty(4\pi t)^{n/2}}e^{-\frac{1}{4t} (h^{-1}x -x)^2} \qty( 1+ \order{x}+ \order{t} ) +\order{t} \nonumber \\
     \to & \frac{\delta^n( x)}{\det_\mathcal{N} (h^{-1} -1 )} +\order{t} \quad (t\to 0).
\end{align}
Using $\det_\mathcal{N}(h)=1$, we get the Atiyah-Bott fixed point theorem \eqref{eq:AtiyahBott}.

%\section{The Computation of the Lefschetz number by Fujikawa Method}
%簡単な状況でFujikwaの方法での計算
%一般的な状況での表式の紹介

%\section{Example}

%

\section{Setup}
\label{sec:Setup}
In this section, we review a magnetized torus model and modular transformation. 

\subsection{Magnetized $T^{2n}$ Model}
\label{subsec:Magetized T2n model}

In this paper, we consider a $U(1)$ gauge theory on the $2n$-dimensional torus $T^{2n}$ and construct the Dirac operator. $T^{2n}$ is given as a quotient manifold of $\mathbb{C}^n \simeq \mathbb{R}^{2n}$. We take $2n$ independent vectors $e_{i},~f_i~(i=1,\cdots,n)$ on the real vector space $\mathbb{R}^{2n}$. We can choose the $n$ vectors $e_i$ which are linearly independent of each other on $\mathbb{C}^n$. Then, there exists an $n\times n$ complex matrix $\Omega=(\Omega_{ij})$ such that 
\begin{align}
    f_j= e_i \Omega_{ij}.
\end{align}
Here, we call $\Omega$ the complex structure of the torus and suppose that $\Omega$ is symmetric and the imaginary part $\Im \Omega$ is positive definite \cite{mumford2007tata}% \footnote{We should impose these assumptions to consider the modular transformation.}
. 
These vectors span a lattice space
\begin{align}
    \Lambda= \Set{ \sum_{i=1}^{n} ( e_i m^i+  f_i n^i)  | m^i,~n^i \in \mathbb{Z}}=\Set{ \sum_{i=1}^{n}  e_i( m^i+  \Omega_{ij} n^j)  | m^i,~n^i \in \mathbb{Z}}.
\end{align}
For simplicity, we let
\begin{align}
    e_i={}^t (0, \cdots, 0,\underbrace{1}_{ \text{i-th spot}},0, \cdots, 0),
\end{align}
where the symbol ${}^t$ means the transpose. We introduce real coordinates 
\begin{align}
    \theta^\mu= \left\{
\begin{array}{ll}
x^i & (\mu=i) \\
y^j & (\mu=n+j)
\end{array}
\right.
\end{align}
along the lattice vectors as
\begin{align}
    \vec{z}=\vec{x} + \Omega \vec{y} 
\end{align}
where $\vec{z}={}^t(z^1,\cdots ,z^n)$ are complex coordinates of $\mathbb{C}^n$. When $\Omega=i 1_n$, $\vec{x}$ and $\vec{y}$ denote the real part and the imaginary part of $\vec{z}$, respectively. 
%We call $\vec{x}$ the real part of $\vec{z}$ and $\vec{y}$ the imaginary part of that, respectively.\footnote{When $\Omega=i 1_n$, this definition is consistent with the ordinary one.} 
%Since $\mathbb{C}^n$ is a complex vector space, a scalar product of the imaginary unit induces a complex structure $J$ on $\mathbb{R}^{2n}$. 
Here, we identified $x^i+1 \sim x^i$ and $y^i +1 \sim y^i$ for all directions and the $2n$-dimensional torus $(T^{2n},\Omega)=\mathbb{C}^n/ \Lambda$ is constructed.

%We fix $\Omega$ and consider a doublet $(T^{2n},\Omega)$. 
The metric on $(T^{2n},\Omega)$ is
\begin{align}
    g=&  \sum_{i,j=1}^n  d (z^i)^\ast  (\Im \Omega)^{-1}_{ij} dz^j \nonumber \\
    =& \mqty({}^td\vec{x} & {}^t d\vec{y})\mqty( 1_n & 0 \\ \Re \Omega & \Im \Omega ) \mqty(\Im \Omega^{-1} & 0 \\ 0 & \Im \Omega^{-1}) \mqty( 1_n & \Re \Omega \\ 0 & \Im \Omega ) \mqty(d\vec{x}\\ d\vec{y}) \nonumber \\
    =&\mqty({}^td\vec{x} & {}^t d\vec{y})\mqty( \Im \Omega^{-1} & \Im \Omega^{-1} \Re \Omega \\ \Re \Omega   \Im \Omega^{-1} & \Re \Omega   \Im \Omega^{-1} \Re \Omega  + \Im \Omega ) \mqty(d\vec{x}\\ d\vec{y}) \label{eq:metric}.
\end{align}
This is different from the Euclidean metric considered in \cite{Kikuchi:2022lfv,Kikuchi:2022psj}. As we will see later, this metric has a good property under the modular transformation. Since the \Lef number and orbifold index do not depend on the metric, we obtain the same result as in \cite{Sakamoto:2020pev,Kobayashi:2022tti,Imai:2022bke,Kikuchi:2022lfv,Kikuchi:2022psj}.

Since $\Im \Omega$ is a positive definite and symmetric matrix, $\Im \Omega$ has a principal square root $\sqrt{\Im \Omega}$. Then, we obtain a vielbein $(e^a_{\mu})$ and the inverse matrix $(E^\mu_a)$:
\begin{align}
    e(\Omega)^a_{\mu}&=  \begin{array}{l cc rr}
 &   \multicolumn{2}{l}{ {\tikz \draw[->] (0,0)-- node[above]{$\mu$} (4,0);} } &&\\
\ldelim({2}{7pt}[] & \sqrt{\Im \Omega}^{-1} & \sqrt{\Im \Omega}^{-1} \Re \Omega & \rdelim){2}{7pt}[]  &\rdelim.{2}{14pt}[{\tikz \draw[<-] (0,0)-- node[right]{$a$} (0,1);}] \\ 
&0 & \sqrt{\Im \Omega}& &\\
&&&&
    \end{array}, \\
    E(\Omega)^\mu_a&= \begin{array}{l cc rr}
 &   \multicolumn{2}{l}{{\tikz \draw[->] (0,0)-- node[above]{$a$} (4,0);} } &&\\
\ldelim({2}{7pt}[] & \sqrt{\Im \Omega} &  -\Re \Omega \sqrt{\Im \Omega}^{-1} & \rdelim){2}{7pt}[]  &\rdelim.{2}{14pt}[{\tikz \draw[<-] (0,0)-- node[right]{$\mu$} (0,1);}] \\ 
&0 & \sqrt{\Im \Omega}^{-1}  & &\\
&&&&
    \end{array}.
\end{align}

This metric leads to the volume form 
\begin{align}
    \text{vol}= \frac{1}{\det(\Im \Omega)} \prod_{j=1}^n \qty(\frac{1}{2i} (dz^j)^\ast \wedge d{z}^{j})=  \prod_{j=1}^n \qty(  dx^j \wedge d{y}^j)
\end{align}
on $T^{2n}$. 
%Here, we let $C= (\det( \Im \Omega)  )^{\frac{1}{n}}$ and normalize the volume of $T^{2n}$. 
Note that the orientation does not always coincide with $dx^1 \wedge \cdots \wedge dx^n \wedge dy^1 \wedge \cdots \wedge dy^n  $.
%We assume that the volume of $T^{2n}$ is one, then we get $C= (\det( \Im \Omega)  )^{\frac{1}{n}}$.

On the torus, we consider a Dirac fermion $\psi$ with pseudo-periodic boundary conditions
\begin{align}
%\begin{aligned}
\psi( \vec{z}+e_k,\Omega )&= \exp( i\pi \qty(  p^x_{kj}x^j + {}^t(N)_{kj}y^j +2 \alpha_k ))\psi(\vec{z},\Omega )=\exp(i\pi \Lambda_k^x )\psi(\vec{z},\Omega ), \label{eq:pseudo-PBC1}\\
    \psi(\vec{z}+\Omega e_k,\Omega )&= \exp( i\pi \qty(  p^y_{kj}y^j -N_{kj}x^j +2 \beta_k ))\psi(\vec{z},\Omega )=\exp(i\pi \Lambda_k^y )\psi(\vec{z},\Omega )\label{eq:pseudo-PBC2},
%\end{aligned}
\end{align}
where $p^x_{kj},~p^y_{kj},~N_{kj}$ are integers, and $\alpha_k$ and $\beta_k$ are real numbers called the Scherk-Schwarz
(SS) phases \cite{Scherk:1978ta}. In general,
\begin{align}
    \psi(\vec{z}+\vec{m}+\Omega \vec{n},\Omega)=&\exp(i\pi \qty(\sum_{i<j} (p^x_{ij} m_i m_j +p^y n_i n_j)+{}^t\vec{m} {}^t N \vec{n} ))\exp(i\pi (m_k \Lambda_k^x +n_k \Lambda_k^y ))\psi(\vec{z},\Omega  ) \nonumber \\
    =&\exp(i\pi\chi( \vec{z}+\vec{m}+\Omega \vec{n};\vec{z},\Omega))\psi(\vec{z},\Omega ).
\end{align}

The covariant derivative for $\psi$ must satisfy the same boundary conditions. Let $A$ be a $1$-form gauge connection. Then, the (external) covariant derivative is defined by
\begin{align}
    (d-iA)\psi(\vec{z},\Omega ).
\end{align}
From the assumption, the boundary conditions for $A$ are given by
\begin{align}
    A(\vec{z}+e_k,\Omega)= A(\vec{z},\Omega )+\pi d\Lambda_k^x (\vec{z},\Omega ) ,\\
    A(\vec{z} + \Omega e_k,\Omega )= A(\vec{z},\Omega )+\pi d\Lambda_k^y (\vec{z},\Omega ).
\end{align}
Thus, we can take
\begin{align}
    A(\vec{z},\Omega )=& \pi \qty(  p^x_{kj} x^k dx^j +  {}^t(N)_{kj} \qty( x^k dy^j - y^j dx^k )+  p^y_{kj} y^k dy^j  )\nonumber \\
    =&\pi \mqty({}^t \vec{x} & {}^t \vec{y} )  \mqty( p^x & {}^tN \\-N & p^y ) \mqty(d\vec{x} \\  d\vec{y} )
\end{align}
which leads to a uniform flux written as
\begin{align}
    F(\vec{z},\Omega )=&dA= \pi\qty( p^x_{ij} dx^{i} \wedge dx^j + 2 {}^tN_{ij} dx^{i} \wedge dy^j + p^y_{ij} dy^{i} \wedge dy^j) \nonumber \\
    =& \pi \mqty({}^t d\vec{x} & {}^t d\vec{y} ) \wedge \mqty( p^x & {}^tN \\-N & p^y ) \mqty(d\vec{x} \\  d\vec{y} ).
\end{align}
Without loss of generality, we suppose that the matrices $p^x=(p^x_{ij})$ and $p^y=(p^y_{ij})$ are anti-symmetric.

Let $\gamma^i~(i=1,\cdots,2n)$ be gamma matrices with $\qty{\gamma^i, \gamma^j}=2 \delta^{ij}$. The gamma matrices along the lattice are written as
\begin{align}
    \Gamma^{x^i}_\Omega&=  \sqrt{\Im \Omega}^{ij} \gamma^{2j-1} -   (\Re \Omega \sqrt{\Im \Omega}^{-1})^{ij} \gamma^{2j},\\
    \Gamma^{y^i}_\Omega&=   (\sqrt{\Im \Omega}^{-1})^{ij} \gamma^{2j}.   
\end{align}
Note that the order of $\gamma^i$ respects the orientation of $T^{2n}$. The Dirac operator with the $U(1)$ gauge field is obtained by
\begin{align}
    \slashed{D}_\Omega= \Gamma^\mu_\Omega \qty( \pdv{}{\theta^\mu} - i A_\mu)=\Gamma^{x^i}_\Omega \qty( \pdv{}{x^i} - i A_{x^i})+\Gamma^{y^i}_\Omega \qty( \pdv{}{y^i} - i A_{y^i})
\end{align}
and the chiral operator is defined by
\begin{align}
    \bar{\gamma}= (-i \gamma^1 \gamma^2)\cdots (-i \gamma^{2n-1} \gamma^{2n}).
\end{align}

\begin{comment}
    \begin{align}
    F= 2\pi \qty( p_x dx^1 \wedge dx^2 + p_y dy^1 \wedge dy^2 + \mqty(dx^1, dx^2) {}^t N \mqty(dy^1 \\dy^2))
\end{align}
\end{comment}

\subsection{Modular Transformation}
In this subsection, we review the modular transformation on the torus $T^{2n} \simeq\mathbb{C}^n / \Lambda$. We fix $\Omega$ and begin with the transformation on $(T^{2n},\Omega)$, which is given by the symplectic group $Sp(2n,\mathbb{Z})$. This leads to the action on the Dirac spinor, the $U(1)$ gauge connection, and the SS phases.

%which preserves the lattice $\Lambda$. This transformation is 
The symplectic group is defined as
\begin{align}
    Sp(2n,\mathbb{Z})= \Set{ h \in GL(2n , \mathbb{Z})|   {}^t h \mqty( 0  & -1_n \\ 1_n & 0) h=   \mqty( 0  & -1_n \\ 1_n & 0) }.
\end{align}
The matrix element $h$ is represented as
\begin{align}
    h= \mqty(A & B \\ C & D)
\end{align}
by $n \times n$ matrices $A,~B,~C$, and $D$ such that
\begin{align}
    {}^t A C= {}^t C A,~ {}^t B D= {}^t D B,~{}^t AD- {}^t C B=1_n. \label{eq:symplectic conditions}
\end{align}
Note that $\det(h)=1$, and the inverse matrix of $h$ is 
\begin{align}
    h^{-1}= \mqty({}^tD & -{}^t B \\- {}^tC & {}^t A).
\end{align}
The generators of $Sp(2n,\mathbb{Z})$ are given by
\begin{align}
    S= \mqty( 0  & -1_n \\ 1_n & 0),~T(B)= \mqty( 1_n  & B \\  0 & 1_n),~ U(A)=\mqty( A & 0 \\ 0 & {}^tA^{-1}),
\end{align}
where $B$ is a symmetric matrix and $A$ belongs to $GL(n, \mathbb{Z})$. 

%In order to construct an orbifold from $T^{2n}$, we focus on a cyclic subgroup $H$ in $Sp(2n,\mathbb{Z})$. 

We fix $\Omega$. The element $h \in Sp(2n, \mathbb{Z})$ induces a map $(T^{2n}, \Omega) \to (T^{2n}, \Omega^\prime)$ defined as
\begin{align}
    \Omega \to  \Omega^\prime= (A\Omega+B)(C\Omega+D)^{-1}=h(\Omega)
\end{align}
and
\begin{align}
    \vec{z}\to \vec{z}^\prime= {}^t( C\Omega +D )^{-1}\vec{z}=  A \vec{x}-B \vec{y }+ \Omega^\prime ( -C \Vec{x} +D \vec{y})= \vec{x}^\prime+ \Omega^\prime  \vec{y}^\prime.
\end{align}
This transformation leads to
\begin{align}
    \mqty(\vec{x} \\ \vec{y}) \to  \mqty(\vec{x}^\prime \\ \vec{y}^\prime)= \bar{h}\mqty(\vec{x} \\ \vec{y})  =\mqty(A & -B \\ -C & D) \mqty(\vec{x} \\ \vec{y}) 
\end{align}
for the real coordinates. Since the imaginary part of $\Omega$ turns into
\begin{align}
    \Im \Omega \to \Im \Omega^\prime ={}^t(C \Omega+D)^{-1} \Im \Omega (C \Omega^\ast+D)^{-1},
\end{align}
the metric \eqref{eq:metric} is invariant under the modular transformation:
\begin{align}
    ds^2 \to ds^{2\prime} = {}^t d \vec{z}^{\prime \ast} \Im \Omega^{\prime -1} d\vec{z}^\prime={}^t d \vec{z}^\ast \Im \Omega^{-1} d\vec{z}=ds^2.
\end{align}

We consider an $n\times n$ complex matrix
\begin{align}
    U(h,\Omega)= \sqrt{\Im h(\Omega)}^{-1} {}^t (C\Omega +D)^{-1} \sqrt{\Im \Omega} .
\end{align}
This matrix satisfies
\begin{align}
    U(g, h(\Omega))U(h,\Omega)=U(gh ,\Omega),~U(h,\Omega)^\dagger=U(h,\Omega)^{-1}
\end{align}
for all $g,h \in Sp(2n ,\mathbb{Z})$, so $U(h,\Omega)$ is a unitary representation of $h$ at fixed $\Omega$. We denote the realification of $U(h,\Omega)=\Re U +i \Im U$ by $U_\mathbb{R}(h,\Omega)=\mqty( \Re U & - \Im U \\ \Im U & \Re U)$. Then, the vielbein changes 
\begin{align}
     e(\Omega) ^a_\mu \to e(\Omega^\prime)^a_\mu = U_\mathbb{R}(h,\Omega)^a_b e(\Omega)^b_\nu (\bar{h}^{-1})^\nu_\mu  \label{eq:action on vielbein}. 
\end{align}

%Under this transformation, a Dirac spinor $\psi_\Omega$ and gamma matrices $\Gamma_\Omega^\mu$ on $(T^{2n},\Omega)$ changes to 
The action on the Dirac spinor $\psi$ on $(T^{2n},\Omega)$ is given by
\begin{align}
   (h\psi)(\vec{z}^\prime,\Omega^\prime)= \rho(h,\Omega) \psi(h^{-1}(\vec{z}^\prime,\Omega^\prime)) \label{eq:action on spinor},
   %h\Gamma_\Omega^\mu=& \Gamma_{\Omega^\prime}^\mu,%\rho(h,\vec{z}^\prime) \Gamma_{\Omega^\prime}^\mu \rho(h,\vec{z}^\prime)^{-1}
\end{align}
where $\rho(h,\Omega)$ denotes a spin$^c$ representation\footnote{The spin$^c$ representation is a combination of $U(1)$ and spin representations.} of $U_\mathbb{R}(h,\Omega)$ such that
\begin{align}
    \rho(h,\Omega)\gamma^a \rho(h,\Omega)^{-1}&= U_\mathbb{R}^{-1}(h,\Omega)^a_b \gamma^b, \label{eq:property of rho(h)}\\
    \rho(g, h(\Omega)) \rho(h,\Omega)&=\rho(gh, \Omega)
\end{align}
for any $g,h \in Sp(2n ,\mathbb{Z})$. 
We briefly explain the construction of $\rho(h,\Omega)$ proposed in Ref. \cite{Atiyah:1964zzClifford}. Let $u_1,\cdots u_n \in \mathbb{C}^n$ be an orthogonal basis such that $U(h,\Omega) u_r= e^{i\theta_r} u_r$. We denote the realification of $u_r$ and $iu_r$ by $v_{2r-1}=\mqty(\Re u_r \\ \Im u_r)$ and $v_{2r}=\mqty(-\Im u_r \\ \Re u_r)$, respectively. Let $P$ be a $2n \times 2n$ orthogonal matrix whose $i$-th column is the eigenvector $v_i$. Then, we have
\begin{align}
    U_\mathbb{R}(h,\Omega) P= P \mqty( \cos \theta_1 & \multicolumn{1}{c|}{-\sin \theta_1} &  &  &  \\ 
\sin \theta_1 & \multicolumn{1}{c|}{\cos \theta_1} &  &  &  \\ \cline{1-4}
 &  & \multicolumn{1}{|c}{\cos \theta_2} & \multicolumn{1}{c|}{-\sin \theta_2} &  \\ 
 &  & \multicolumn{1}{|c}{\sin \theta_2} & \multicolumn{1}{c|}{\cos \theta_2} &  \\ \cline{3-5}
 &  &  &  & \multicolumn{1}{|c}{\ddots} \\ )=PD.
\end{align}
We put $\gamma^{\prime a}= (P^{-1})^a_b \gamma^b$ and obtain
\begin{align}
    \rho(h,\Omega)= \prod_{r=1}^n \qty(\cos \frac{\theta_r}{2} - \gamma^{\prime 2r-1} \gamma^{\prime 2r}\sin \frac{\theta_r}{2}) e^{i\frac{\theta_r}{2}}.
\end{align}
According to the equation $\eqref{eq:action on vielbein}$, the $\Gamma_\Omega^{x^i}$ and $\Gamma_\Omega^{y^i}$ change as
\begin{align}
    \rho(h) \Gamma_\Omega^{x^i} \rho(h)^{-1}&= {}^tD_{ia} \Gamma^{x^{\prime a}}_{\Omega^\prime} + {}^tB_{ia} \Gamma^{y^{\prime a}}_{\Omega^\prime} ,\\
    \rho(h) \Gamma_\Omega^{y^i} \rho(h)^{-1}&= {}^tC_{ia} \Gamma^{x^{\prime a}}_{\Omega^\prime} + {}^tA_{ia} \Gamma^{y^{\prime a}}_{\Omega^\prime}.
\end{align}

In this construction, the ambiguity of the $U(1)$ phase factor remains. We can alter the spin$^c$ representation as
\begin{align}
    \rho(h,\Omega) \to \rho^\prime(h,\Omega)= \rho(h,\Omega) \det(U(h,\Omega) )%\frac{\det({}^t(C\Omega+D)^{-1})}{ \abs{\det( {}^t(C\Omega+D)^{-1})}},
\end{align}
where the determinant denotes one of the $U(1)$ representations. In this paper, we determine the phase factor as follows. We first define the Clifford conjugate of an element $x$ in Clifford algebra by $\bar{x}=\bar{\gamma} x^T \bar{\gamma}$, where $x^T$ reverses the order of gamma matrices in each term. For example, the conjugate of $x=\gamma^1+ \gamma^1 \gamma^2 + \gamma^1 \gamma^2 \gamma^3 \gamma^4$ is
\begin{align}
    \bar{x} =& -\gamma^1+ \gamma^2 \gamma^1 + \gamma^4 \gamma^3 \gamma^2 \gamma^1 =-\gamma^1- \gamma^1 \gamma^2 + \gamma^1 \gamma^2 \gamma^3 \gamma^4.
\end{align}
Then, we choose the phase factor as
\begin{align}
\overline{\rho(h,\Omega)} \rho(h,\Omega)=\det(U(h,\Omega)).
\end{align}

We assume that the $U(1)$ gauge connection transforms as
\begin{align}
    A_{x^{\prime a}}^\prime(\vec{z}^\prime,\Omega^\prime)=&  A_{x^{i}}(\vec{z},\Omega) {}^tD_{ia} + A_{y^{i}}(\vec{z},\Omega) {}^tC_{ia},\\
    A_{y^{\prime a}}^\prime (\vec{z}^\prime,\Omega^\prime)=& A_{x^i}(\vec{z},\Omega) ~{}^t B_{ia}  +A_{y^i}(\vec{z},\Omega) ~{}^t A_{ia}.
\end{align}
These transformation laws for the $U(1)$ connections are equivalent to
\begin{align}
     \mqty( p^{x \prime } & {}^tN^\prime \\-N^\prime & p^{y \prime } ) = {}^t \bar{h}^{-1}\mqty( p^x & {}^tN \\-N & p^y ) \bar{h}^{-1} \label{eq:action on U(1)} .
\end{align}
Then, the Dirac operator commutes with the transformation $h$:
\begin{align}
    h \slashed{D}_\Omega= \slashed{D}_{\Omega^\prime} h.
\end{align}

\begin{comment}

\begin{align}
    \rho(g, h(\Omega)) \rho(h,\Omega)=\rho(gh, \Omega)
\end{align}
for any $g$ and $h$. We assume $h \slashed{D}_\Omega= \slashed{D}_{\Omega^\prime} h $, then we get 
\begin{align}
    \rho(h) \Gamma_\Omega^{x^i} \rho(h)^{-1}= {}^tD_{ia} \Gamma^{x^{\prime a}}_{\Omega^\prime} + {}^tB_{ia} \Gamma^{y^{\prime a}}_{\Omega^\prime} \\
    \rho(h) \Gamma_\Omega^{y^i} \rho(h)^{-1}= {}^tC_{ia} \Gamma^{x^{\prime a}}_{\Omega^\prime} + {}^tA_{ia} \Gamma^{y^{\prime a}}_{\Omega^\prime}
\end{align}
and
\begin{align}
    A_{x^{\prime a}}^\prime(\vec{z}^\prime,\Omega^\prime)=&  A_{x^{i}}(\vec{z},\Omega) {}^tD_{ia} + A_{y^{i}}(\vec{z},\Omega) {}^tC_{ia},\\
    A_{y^{\prime a}}^\prime (\vec{z}^\prime,\Omega^\prime)=& A_{x^i}(\vec{z},\Omega) ~{}^t B_{ia}  +A_{y^i}(\vec{z},\Omega) ~{}^t A_{ia}.
\end{align}
These transformation lows for the $U(1)$ connections are equivalent to
\begin{align}
     \mqty( p^{x \prime } & {}^tN^\prime \\-N^\prime & p^{y \prime } ) = {}^t \bar{h}^{-1}\mqty( p^x & {}^tN \\-N & p^y ) \bar{h}^{-1} \quad \qty(\bar{h}=\mqty( A & -B \\ -C & D) ) \label{eq:action on U(1)} .
\end{align}
\end{comment}

Furthermore, $h\psi$ must satisfy the pseudo-periodic boundary conditions \eqref{eq:pseudo-PBC1} and \eqref{eq:pseudo-PBC2}. It implies that the SS phases change
\begin{align}
    \exp(i\pi 2\alpha_k^\prime )=& \exp( i\pi 2 \qty(D_{ki}\alpha_i + C_{ki}\beta_i ) ) \exp( i\pi \qty( \sum_{i<j}(p^x_{ij} D_{ki}D_{kj} + p^y_{ij} C_{ki}C_{kj}) +{}^t N_{ij} D_{ki} C_{kj}  )), \label{eq:action on SS1} 
    \\
    \exp(i\pi 2\beta_k^\prime )=& \exp( i\pi 2 \qty(B_{ki}\alpha_i + A_{ki}\beta_i ) ) \exp( i\pi \qty( \sum_{i<j}(p^x_{ij} B_{ki}B_{kj} + p^y_{ij} A_{ki}A_{kj}) +{}^t N_{ij} B_{ki} A_{kj}  )) .  \label{eq:action on SS2} 
\end{align}

\subsection{Magnetized $T^{2n}/ \mathbb{Z}_N$ model}
\begin{comment}
If a cyclic subgourp $H\simeq \mathbb{Z}_N$ does not change the lattice $\Lambda$, it can be extended to the torus $\mathbb{C}^n / \Lambda$. Then, we get an orbifold $T^{2n}/ \mathbb{Z}_N $. As a result, the $U(1)$ connection and the SS phases must satisfy some conditions.

\end{comment}

In order to construct an orbifold $T^{2n}/\mathbb{Z}_N$ from $(T^{2n},\Omega)$, we focus on a cyclic subgroup $H\simeq \mathbb{Z}_N$ in $Sp(2n,\mathbb{Z})$. For example, $Sp(2,\mathbb{Z})$ includes subgroups that are equivalent to $\mathbb{Z}_2,~\mathbb{Z}_3,~\mathbb{Z}_4$, and $\mathbb{Z}_6$. Let $h$ be a generator of $H$ such that
\begin{align}
    h(\Omega)=\Omega.
\end{align}
Then, $H$ turns into a $\mathbb{Z}_N$-transformation on $T^{2n}$, and we get an orbifold $T^{2n}/\mathbb{Z}_N$. The action on the spinor is given by \eqref{eq:action on spinor} replaced ${\Omega^\prime}$ with ${\Omega}$. We also obtain $H$-invariant conditions by eliminating $\prime$ in \eqref{eq:action on U(1)}, \eqref{eq:action on SS1}, and \eqref{eq:action on SS2}.

A fixed point $\vec{z}_f$ of the transformation $h$ is given by
\begin{align}
    {}^t ( C\Omega + D) \vec{z}_f= \vec{z}_f+ \vec{m}+\Omega \vec{n},
\end{align}
where $\vec{m}+\Omega \vec{n} \in \Lambda$ and $\vec{z}_f = \vec{x}_f+\Omega \vec{y}_f$ should be $0\leq x_f^i,y_f^i <1$. At the fixed point, $h$ acts as
\begin{align}
    h\psi(\vec{z}_f,\Omega)= \rho(h,\Omega) \psi( {}^t ( C\Omega + D)\vec{z}_f, \Omega )= & \rho(h,\Omega) \exp(i\pi \chi( {}^t ( C\Omega + D)\vec{z}_f ;\vec{z}_f,\Omega)) \psi(\vec{z}_f,\Omega),\\
    \rho(h,z_f)= & \rho(h,\Omega) \exp(i\pi \chi( {}^t ( C\Omega + D)\vec{z}_f ;\vec{z}_f,\Omega)) .
\end{align}
Then, the \Lef number of $h$ is given by Eq.~\eqref{eq:AtiyahSegalSinger} and \eqref{eq:AtiyahBott}, and the index of $h\psi= e^{i\frac{2\pi k}{N}} \psi$ sector is obtained by
\begin{align}
    \text{ind}_{T^{2n}/\mathbb{Z}_N}(h= e^{i\frac{2\pi k}{N}})= \frac{1}{N}\sum_{l=0}^{N-1} \qty( L(h^l)  e^{-i\frac{2\pi kl}{N}} )  \label{eq:formula_of_index}. 
\end{align}

For any $g \in Sp(2n, \mathbb{Z})$, $gHg^{-1}$ is also a cyclic subgroup of order $N$ generated by $g h g^{-1}$. The transformation $g$ is a map $(T^{2n}, \Omega)\to (T^{2n}, g(\Omega))$, and  $g h g^{-1}$ becomes the $\mathbb{Z}_N$-transformation on $(T^{2n}, g(\Omega))$. Here, the $U(1)$ gauge connection and SS phases change as in \eqref{eq:action on U(1)}, \eqref{eq:action on SS1}, and \eqref{eq:action on SS2}. On the other hand, the \Lef number and index do not change under the transformation $g$. It implies that we only need to consider the representative element of the conjugacy class.

\section{$T^2/\mathbb{Z}_N~(N=2,3,4,6)$}
\label{T2ZN}
In this section, we consider the orbifolds $T^2/\mathbb{Z}_N~(N=2,3,4,6)$ and compute the orbifold index by using the fixed point theorem. In the two-dimensional theory, the invariant flux is written as
\begin{align}
    F=2\pi M dx \wedge dy~(M\in \mathbb{Z}).
\end{align}
and the chiral operator is
\begin{align}
    \bar{\gamma}=\sigma_3,
\end{align}
where $\gamma^1=\sigma_1$ and $\gamma^2=\sigma_2$.

\subsection{ $T^2/ \mathbb{Z}_2$}
\label{subsec:T2Z2}

$T^2/\mathbb{Z}_2$ is established by
\begin{align}
    h=\mqty(-1 & 0 \\ 0& -1).
\end{align}
This acts on $\Omega$ trivially, so any $\Omega$ satisfies $h(\Omega)=\Omega$. Since the unitary representation is $-1$, the action on the Dirac spinor is given by
\begin{align}
    h\psi(z,\Omega)= \sigma_3 \psi(-z,\Omega).
\end{align}

According to equations \eqref{eq:action on SS1} and \eqref{eq:action on SS2}, the SS phases are quantized as \cite{Abe:2013bca} %SS phases are regulated by $\exp(i\pi 2 \alpha)= \exp(-i\pi 2 \alpha)$ so we get
\begin{align}
     (\alpha ,\beta )=(0,0),~\qty(0,\frac{1}{2}),~\qty(\frac{1}{2},0),~\qty(\frac{1}{2},\frac{1}{2}). 
\end{align}

Under this transformation, there are four fixed points:
\begin{align}
    z_f=\frac{a +\Omega b}{2} \quad (a,b=0,1).
\end{align}
At the fixed points, $h$ acts as
\begin{align}
    \rho(h, z_f)= \sigma_3 \exp( i \pi \mqty(a & 1) \mqty( M & -2 \alpha \\ -2 \beta & 0 ) \mqty(b \\ 1)).
\end{align}
Thus, the \Lef numbers are given by
\begin{align} 
L(1)=& M, \\
    L(-1)= &\sum_{a,b=0,1}\frac{ \tr(\sigma_3 \rho(h, z_i^f)) }{4}
    =\frac{1}{2}\qty[1+ e^{-i2\pi\alpha_1} +e^{-i2\pi\alpha_2}+e^{i\pi M }e^{-i2\pi\alpha_1}e^{-i2\pi\alpha_2}],
\end{align}
and the indices are obtained by
\begin{align}
\text{ind}_{T^2/{\mathbb{Z}_2}}(h=1 )=& \frac{1}{2}(L(1)+L(e^{i\pi})) = \qty[\frac{M+1}{2}] + e^{-i\pi M} \frac{1+ e^{-i2\pi\alpha_1} e^{i\pi M} }{2}\frac{1+ e^{-i2\pi\alpha_2} e^{i\pi M} }{2}, \\
\text{ind}_{T^2/{\mathbb{Z}_2}}(h=-1 )=& \frac{1}{2}(L(1)-L(e^{i\pi})) = \qty[\frac{M}{2}] - e^{-i\pi M} \frac{1+ e^{-i2\pi\alpha_1} e^{i\pi M} }{2}\frac{1+ e^{-i2\pi\alpha_2} e^{i\pi M} }{2},
\end{align}
where $[\ast]$ denotes the Gauss symbol, which takes the integer part of a given real number. These results are consistent with the previous works \cite{Sakamoto:2020pev,Kobayashi:2022tti,Imai:2022bke} and make it clear that the orbifold index is an integer.

\subsection{ $T^2/ \mathbb{Z}_3$}

The element $h$ with $h^3=1$ belongs to a conjugacy class of the following elements:
\begin{align}
     \mqty(0 & -1 \\ 1 & -1),~\mqty(0 & +1 \\ -1 & -1).
\end{align}
Without loss of generality, we consider $T^2/ \mathbb{Z}_3$ constructed by $h=\mqty(0 & -1 \\ 1 & -1)$, since $h^2 $ is conjugate to $\mqty(0 & +1 \\ -1 & -1)$. The invariant complex structure $\Omega$ under $h$ is given by
\begin{align}
   \Omega= \frac{1}{2}( 1+ i\sqrt{3})=e^{i\frac{\pi}{3}},
\end{align}
and the action on the Dirac spinor is obtained by
\begin{align}
    h\psi(z,\Omega)= \exp(\frac{\pi}{3} \sigma_1 \sigma_2)e^{-i\frac{\pi}{3}} \psi( e^{i\frac{2\pi}{3}}z,\Omega).
\end{align}

The consistent SS phases must satisfy
\begin{align}
    e^{i\pi 2 \alpha}= e^{i\pi 2 ( - \alpha + \beta)} e^{-i\pi M},~e^{i\pi 2 \beta}= e^{i\pi 2 (-\alpha)},
\end{align}
and we have $\alpha +\beta \in \mathbb{Z},~ 6\beta-M \in 2\mathbb{Z}$. Then, we choose these parameters as follows:
\begin{align}
    \alpha =- \beta= -\frac{M}{6} + \frac{u}{3} \quad (u\in \mathbb{Z}).
\end{align}

We find three fixed points under this transformation:
\begin{align}
    z^f= \frac{c}{3}(1+ \Omega) \quad (c \in \mathbb{Z}).
\end{align}
At the fixed points, $h$ acts as
\begin{align}
    \rho(h,z_f)=\exp(\frac{\pi}{3} \sigma_1 \sigma_2)e^{-i\frac{\pi}{3}}\exp(i \pi \left(- \frac{M c^{2}}{3} + c \left(\frac{M}{3} - \frac{2 u}{3}\right)\right)),
\end{align}
so the \Lef numbers are given by
\begin{align}
    L(1)&=M ,\\
    L(\omega)&= \frac{1}{3}(1-\omega^2) ( 1+\omega^{2u}+ \omega^{u+2M} ), \\
    L(\omega^2)&=\frac{1}{3}(1-\omega) (1+\omega^{u}+ \omega^{2u+M}),
\end{align}
where $\omega=e^{i\frac{2\pi}{3}}$.  
Then, we have the indices:
\begin{align}
    \text{ind}_{T^2/{\mathbb{Z}_3}}(h=1 )=& \frac{1}{3}(L(1)+L(\omega)+L(\omega^2) ) \nonumber \\
    =&\frac{1}{3}\qty( M+1 +\frac{2}{\sqrt{3}} \sin(2\pi \frac{M+1}{3}) 2\cos (4\pi \alpha) )\nonumber \\
    =& \qty[ \frac{M+2}{3}] +\frac{2}{\sqrt{3}} \sin(2\pi \frac{M+1}{3}) \frac{1+ 2\cos (2\pi \frac{M-2u}{3})}{3}, \\
    \text{ind}_{T^2/{\mathbb{Z}_3}}({h=\omega })=& \frac{1}{3}(L(1)+ \omega^{-1} L(\omega)+ \omega^{-2} L(\omega^2) ) \nonumber\\
    =&\frac{1}{3}\qty( M +\frac{2}{\sqrt{3}} \sin(2\pi \frac{M}{3}) 2\cos (4\pi \alpha) ) \nonumber \\
    =& \qty[ \frac{M+1}{3}] +\frac{2}{\sqrt{3}} \sin(2\pi \frac{M}{3}) \frac{1+ 2\cos (2\pi \frac{M-2u}{3})}{3},\\
    \text{ind}_{T^2/{\mathbb{Z}_3}}({h=\omega^2 })=& \frac{1}{3}(L(1)+ \omega^{-2} L(\omega)+ \omega^{-4} L(\omega^2) ) \nonumber \\
    =&\frac{1}{3}\qty( M-1 +\frac{2}{\sqrt{3}} \sin(2\pi \frac{M-1}{3}) 2\cos (4\pi \alpha) ) \nonumber \\
    =& \qty[ \frac{M}{3}] +\frac{2}{\sqrt{3}} \sin(2\pi \frac{M-1}{3}) \frac{1+ 2\cos (2\pi \frac{M-2u}{3})}{3}.
\end{align}
These results are consistent with Ref.~\cite{Sakamoto:2020pev,Kobayashi:2022tti,Imai:2022bke}.

\subsection{ $T^2/ \mathbb{Z}_4$}
The element $h$ with $h^4=1$ belongs to a conjugacy class of the following elements:
\begin{align}
    h=\mqty(0 & -1 \\ 1 & 0 ),~h^\prime=\mqty(0 & 1 \\ -1 & 0 )=h^3. 
\end{align}
The invariant complex structure $\Omega$ under $h$ is given by
\begin{align}
    \Omega=i,
\end{align}
and the action on the Dirac spinor is obtained by
\begin{align}
    h\psi(z,\Omega)= \exp(\frac{\pi}{4} \sigma_1 \sigma_2)e^{-i\frac{\pi}{4}} \psi( iz,\Omega).
\end{align}
The consistent SS phases must satisfy
\begin{align}
    e^{i\pi 2 \alpha}= e^{i\pi 2 \beta} ,~e^{i\pi 2 \beta}= e^{i\pi 2 (-\alpha)},
\end{align}
and we have $\alpha=\beta=0,~\frac{1}{2}$.

$h$ with $h^4=1$ has two $\mathbb{Z}_4$ fixed points:
\begin{align}
    z^f_{\mathbb{Z}_4}=\frac{c}{2}(1+i) \quad (c=0,~1).
\end{align}
At the ${\mathbb{Z}_4}$ fixed points, $h$ acts as 
\begin{align}
    \rho(h,z^f_{\mathbb{Z}_4})=\exp(\frac{\pi}{4} \sigma_1 \sigma_2)e^{-i\frac{\pi}{4}} e^{i \pi \left(- \frac{M c^{2}}{2} - 2 \alpha c\right)}.
\end{align}
On the other hand, $h^2=-1$ is a generator of a cyclic group of order $2$. As in the previous section, we have
\begin{align}
    \rho(h,z^f_{\mathbb{Z}_2})=\sigma_3 \exp( i \pi \mqty(a & 1) \mqty( M & -2 \alpha \\ -2 \alpha & 0 ) \mqty(b \\ 1))\quad (a,~b=0,~1)
\end{align}
at the $\mathbb{Z}_2$ fixed points
\begin{align}
    z^f_{\mathbb{Z}_2}=\frac{1}{2}(a+ib).
\end{align}

Then, we have the orbifold indices:
\begin{align}
\text{ind}_{T^2/{\mathbb{Z}_4}}(h=1 )=& \frac{1}{4}(L(1)+L(i) +L(-1) +L(-i)) 
 \nonumber \\
=& \frac{1 + e^{i2\pi \alpha }}{2}  \qty[\frac{M+4}{4}] + \frac{1 - e^{i2\pi \alpha }}{2}  \qty[\frac{M+2}{4}], \\
\text{ind}_{T^2/{\mathbb{Z}_4}}({ h=i})=& \frac{1}{4}(L(1)+iL(i) -L(-1) -iL(-i)) \nonumber \\
=& \frac{1 + e^{i2\pi \alpha }}{2}  \qty[\frac{M+1}{4}] + \frac{1 - e^{i2\pi \alpha }}{2}  \qty[\frac{M+3}{4}],\\
\text{ind}_{T^2/{\mathbb{Z}_4}}(h=-1 )=& \frac{1}{4}(L(1)-L(i) +L(-1) -L(-i)) \nonumber \\
=&\frac{1 + e^{i2\pi \alpha }}{2}  \qty[\frac{M+2}{4}] + \frac{1 - e^{i2\pi \alpha }}{2}  \qty[\frac{M}{4}],\\
\text{ind}_{T^2/{\mathbb{Z}_4}}({h=-i })=& \frac{1}{4}(L(1)-iL(i) -L(-1) +iL(-i)) \nonumber\\
=& \frac{1 + e^{i2\pi \alpha }}{2}  \qty[\frac{M-1}{4}] + \frac{1 - e^{i2\pi \alpha }}{2}  \qty[\frac{M+1}{4}].
\end{align}
These results are consistent with Ref.~\cite{Sakamoto:2020pev,Kobayashi:2022tti,Imai:2022bke}.
\subsection{ $T^2/ \mathbb{Z}_6$}
$T^2/ \mathbb{Z}_6$ is constructed by
\begin{align}
    h=\mqty(1 & -1 \\ 1 & 0).
\end{align}
The invariant complex structure $\Omega$ under $h$ is 
\begin{align}
    \Omega= \frac{1}{2}(1+i\sqrt{3}),
\end{align}
and the action on the Dirac spinor is obtained by
\begin{align}
    h\psi(z,\Omega)=\exp(\frac{\pi}{6} \sigma_1 \sigma_2)e^{-i\frac{\pi}{6}} \psi( \xi z,\Omega),
\end{align}
where $\xi= e^{i \frac{2\pi}{6}}$. To be consistent, SS phases must satisfy
\begin{align}
    \exp(i2\pi \alpha)= \exp(i2\pi \beta),~\exp(i2\pi \beta)=\exp(i\pi(2 \beta -2 \alpha-M ) ),
\end{align}
and we obtain $\alpha=M/2$.

$h$ with $h^6=1$ has one $\mathbb{Z}_6$ fixed point:
\begin{align}
    z^f_{\mathbb{Z}_6}=0.
\end{align}
On the other hand, since $h^2$ is a generator of $\mathbb{Z}_3$, we have the following $\mathbb{Z}_3$ fixed points:
\begin{align}
    z^f_{\mathbb{Z}_3}=\frac{c}{3}(1+\Omega)\quad (c=0,1,2).
\end{align}
And since $h^3$ is a generator of $\mathbb{Z}_2$, we also have the following $\mathbb{Z}_2$ fixed points:
\begin{align}
    z^f_{\mathbb{Z}_2}=\frac{1}{2}(a+\Omega b)\quad(a,b=0,1).
\end{align}
At the fixed points, $h$ acts as 
\begin{align}
    \rho(h,z^f_{\mathbb{Z}_6})&=\exp(\frac{\pi}{6} \sigma_1 \sigma_2)e^{-i\frac{\pi}{6}}, \\
    \rho(h,z^f_{\mathbb{Z}_3})&=\exp(\frac{\pi}{6} \sigma_1 \sigma_2)e^{-i\frac{\pi}{6}}e^{i \pi \left(- \frac{M c^{2}}{3} - M c\right)},\\
     \rho(h,z^f_{\mathbb{Z}_2})&=\exp(\frac{\pi}{6} \sigma_1 \sigma_2)e^{-i\frac{\pi}{6}}e^{i \pi \left(M a b - M a - M b\right)},
\end{align}
for the $\mathbb{Z}_6$, $\mathbb{Z}_3$, and $\mathbb{Z}_2$ fixed points, respectively.

% \begin{align}
%     1 \\
% e^{i \pi \left(- \frac{M c^{2}}{3} - M c\right)} \\
% e^{i \pi \left(M a b - M a - M b\right)}
% \end{align}
The indices are obtained as
\begin{align}
    \text{ind}_{T^2/{\mathbb{Z}_6}}(h=+1)=& \frac{1}{6}\qty( M+1 +2 \frac{2}{\sqrt{3}} \sin (2\pi \frac{M+1}{3})+ 3 \frac{1+(-1)^M}{2} ),\\
    %=&\qty[ \frac{\text{ind}(h^2=+1) +1}{2} ]
    \text{ind}_{T^2/{\mathbb{Z}_6}}(h=\xi)=& \frac{1}{6}\qty( M +2 \frac{2}{\sqrt{3}} \sin (2\pi \frac{M}{3})+3 \frac{1-(-1)^M}{2} ) ,\\
    \text{ind}_{T^2/{\mathbb{Z}_6}}(h=\xi^2)=& \frac{1}{6}\qty( M -1+ 2 \frac{2}{\sqrt{3}} \sin (2\pi \frac{M-1}{3})+3 \frac{1+(-1)^M}{2} ) ,\\
    \text{ind}_{T^2/{\mathbb{Z}_6}}(h=\xi^3)=& \frac{1}{6}\qty( M+1 +2 \frac{2}{\sqrt{3}} \sin (2\pi \frac{M+1}{3})- 3 \frac{1+(-1)^M}{2} ),\\
    \text{ind}_{T^2/{\mathbb{Z}_6}}(h=\xi^4)=& \frac{1}{6}\qty( M +2 \frac{2}{\sqrt{3}} \sin (2\pi \frac{M}{3})-3 \frac{1-(-1)^M}{2} ) ,\\
    \text{ind}_{T^2/{\mathbb{Z}_6}}(h=\xi^5)=& \frac{1}{6}\qty( M -1+ 2 \frac{2}{\sqrt{3}} \sin (2\pi \frac{M-1}{3})-3 \frac{1+(-1)^M}{2} ) .
\end{align}
These results are consistent with Ref.~\cite{Sakamoto:2020pev,Kobayashi:2022tti,Imai:2022bke}.

\section{$T^4/\mathbb{Z}_N(N=2,3,5)$}
\label{T4ZN}
In this section, we attempt to compute orbifold indices on $T^4/\mathbb{Z}_N(N=2,3,5)$, which are constructed by a cyclic group of order $N$, where $N$ is a prime number.

\subsection{$T^4/\mathbb{Z}_2$}
In this subsection, we explore $T^4/\mathbb{Z}_2$ models. The generators are classified into three types:
\begin{align}
    h_1= \left(\begin{array}{cc|cc}
-1 & 0 & & \\ 
0 & -1 && \\ \cline{1-4}
&&-1 & 0 \\
&& 0 & -1  
\end{array}\right),~h_2=\left(\begin{array}{cc|cc}
1 & 0 & & \\ 
0 & -1 && \\ \cline{1-4}
&&1 & 0 \\
&& 0 & -1  
\end{array}\right),~h_3=\left(\begin{array}{cc|cc}
0 & 1 & & \\ 
1 & 0 && \\ \cline{1-4}
&&0 & 1 \\
&& 1 & 0  
\end{array}\right).
\end{align}
Let $(T^4/\mathbb{Z}_2)_i$ be the orbifold generated by $h_i$. 
While all the fixed points of $h_1$ are isolated points, the fixed points of $h_2$ and $h_3$ form a two-dimensional manifold. The flux $F$ can be represented in the form
\begin{align}
    F= & \pi\qty( p^x dx^{1} \wedge dx^2 + 2 {}^tN_{ij} dx^{i} \wedge dy^j + p^y dy^{1} \wedge dy^2) \nonumber \\
    =& \pi \mqty({}^t d\vec{x} & {}^t d\vec{y} ) \wedge \mqty( -p^x \epsilon & {}^tN \\-N & -p^y\epsilon ) \mqty(d\vec{x} \\  d\vec{y} ),
\end{align}
where $\epsilon=\mqty(0 & -1 \\ 1 & 0)$. 

\subsubsection{ $(T^4/\mathbb{Z}_2)_1$}
\label{subsec:T4Z2_1}

The transformation $h_1$ does not change $\Omega$ and the flux $F$, and it acts on the Dirac spinor as 
\begin{align}
    h_1 \psi(\vec{z},\Omega)=& -\exp(i \frac{\pi}{2} \gamma^1 \gamma^2 )\exp(i \frac{\pi}{2} \gamma^3 \gamma^4 )\psi(-\vec{z},\Omega) \nonumber \\
   =& \bar{\gamma} \psi(-\vec{z},\Omega) .
\end{align} 
The SS phases $\alpha_i$ and $\beta_j$ are quantized as either zero or one-half as in the section \ref{subsec:T2Z2}.

We find $16$ fixed points:
\begin{align}
    \vec{z}^f= \frac{1}{2}(\vec{a}+\Omega \vec{b}),
\end{align}
where $\vec{a}={}^t (a_1,a_2)$ and $ \vec{b}={}^t (b_1,b_2)$ are vectors with elements that are either zero or one. At the fixed points, the transformation $h$ is represented by
\begin{align}
%\exp{i \pi \left(- 2 \alpha_{1} c_{1} - 2 \alpha_{2} c_{2} - 2 \beta_{1} d_{1} - 2 \beta_{2} d_{2} + c_{1} c_{2} p_{x} + c_{1} d_{1} n_{11} + c_{1} d_{2} n_{21} + c_{2} d_{1} n_{12} + c_{2} d_{2} n_{22} + d_{1} d_{2} p_{y}\right)} \\
%=\exp( i\pi \qty( \mqty(a_1 & 1) \mqty( p^x & -2 \alpha_1 \\ -2 \alpha_2 & 0 ) \mqty(a_2 \\ 1) + \mqty(b_1 & 1) \mqty( p^y & -2 \beta_1 \\ -2 \beta_2 & 0 ) \mqty(b_2 \\ 1) + \mqty(a_1 & a_2) {}^tN \mqty(b_1 \\ b_2) )) \\
\rho(h,z^f)=-\exp(i \frac{\pi}{2} \gamma^1 \gamma^2 )\exp(i \frac{\pi}{2} \gamma^3 \gamma^4 )\exp( i\pi \qty( p^x a_1 a_2+p^y b_1 b_2 + \mqty({}^t\vec{a} & 1) \mqty( {}^tN & -2\vec{\alpha} \\-2 {}^t\vec{\beta} & 0) \mqty(\vec{b} \\ 1) )).
\end{align}
Then, we have the \Lef numbers
\begin{align}
    L(1)=&-p^x p^y + \det(N), \\
    L(h)=& \frac{1}{4}\sum_{a_1,a_2,b_1,b_2=0,1} \exp( i\pi \qty( p^x a_1 a_2+p^y b_1 b_2 + \mqty({}^t\vec{a} & 1) \mqty( {}^tN & -2\vec{\alpha} \\-2 {}^t\vec{\beta} & 0) \mqty(\vec{b} \\ 1) )),
\end{align}
and orbifold indices
\begin{align}
    \text{ind}(h=+1)=&\frac{1}{2} (L(1)+ L(h)), \\
    \text{ind}(h=-1)=&\frac{1}{2} (L(1)- L(h)).
\end{align}
Compared to the previous section, it is difficult to simplify these formulas. We have verified that these indices are integers through brute-force calculation and found that they are consistent with those in the previous works \cite{Kikuchi:2022lfv}.

\subsubsection{ $(T^4/\mathbb{Z}_2)_2$}
$(T^4/\mathbb{Z}_2)_2$ is constructed by $h_2=\mqty(\sigma_3 & 0 \\ 0 & \sigma_3)$. The invariant complex structure $\Omega$ must commute with $\sigma_3$:
\begin{align}
    h_2(\Omega)=\sigma_3\Omega \sigma_3=\Omega,
\end{align}
so the form is 
\begin{align}
    \Omega=\mqty(\tau_1 & 0 \\ 0 & \tau_2)
\end{align}
with $\Im \tau_i>0~(i=1,2)$. The transformation $h_2$ acts on the Dirac spinor as
\begin{align}
    h_2 \psi(\vec{z}, \Omega)= \exp( \frac{\pi}{2} \gamma^3 \gamma^4)e^{-i\frac{\pi}{2}} \psi(\sigma_3 \vec{z},\Omega). 
\end{align}
The invariant flux also satisfy
\begin{align}
0=\mqty( -p^x\epsilon & {}^tN \\-N & p^y \epsilon)  - {}^t \bar{h}^{-1}_2\mqty( -p^x \epsilon& {}^tN \\-N &- p^y\epsilon ) \bar{h}^{-1}_2 =  \left(\begin{matrix}0 & 2 p_{x} & 0 & 2 n_{21}\\- 2 p_{x} & 0 & 2 n_{12} & 0\\0 & - 2 n_{12} & 0 & 2 p_{y}\\- 2 n_{21} & 0 & - 2 p_{y} & 0\end{matrix}\right),
\end{align}
and we have
\begin{align}
    F= 2\pi( n_{11} dx^1 \wedge dy^1 +n_{22} dx^2 \wedge dy^2) .
\end{align}
The SS phases $\alpha_2$ and $\beta_2$ are quantized as
\begin{align}
    (\alpha_2,\beta_2)=(0,0),~\qty(0,\frac{1}{2}),~\qty(\frac{1}{2},0),~\qty(\frac{1}{2},\frac{1}{2}). 
\end{align}
However, $\alpha_1$ and $\beta_1$ can take arbitrary real numbers.

In contrast to the previous sections, the fixed points form four two-dimensional tori:
\begin{align}
    \vec{z}^f=\mqty( x^1 \\ a/2)+ \Omega \mqty(y^1 \\ b/2),
\end{align}
where $a,b=0,1$. We denote the manifold defined by this equation as $M^{h_2}$.

The \Lef numbers are given by the Atiyah-Segal-Singer fixed point theorem \eqref{eq:AtiyahSegalSinger}. The normal bundle $\mathcal{N}$ of $M^{h_2}$ is spanned by $\qty(\pdv{}{x^2},\pdv{}{y^2})$ on each point, so the restriction of $h_2$ to $\mathcal{N}$ is represented as $h_2 |_\mathcal{N}= -1_2$, and the determinant is
\begin{align}
    \text{det}_\mathcal{N}(1-h_2 )=2^2=4.
\end{align} 
On $M^{h_2}$, the flux is 
\begin{align}
    F|_{M^{h_2}}= 2\pi n_{11} dx^1 \wedge dy^1. 
\end{align}
The spinor bundle $\mathbf{S}$ can be decomposed as 
\begin{align}
    \mathbf{S}|_{M^{h_2}}=\mathbf{S}^\prime \otimes \mathbf{S}^{\prime \prime} ,
\end{align}
where $\mathbf{S}^\prime$ and $\mathbf{S}^{\prime \prime}$ are spinor bundles associated with $TM^{h_2}$ and $\mathcal{N}$, respectively. Under this decomposition, the gamma matrices are represented as
\begin{align}
    \gamma^1= \sigma^1 \otimes \sigma_3,~ \gamma^2= \sigma^2 \otimes \sigma_3,~ \gamma^3= 1_2 \otimes \sigma_1,~\gamma^4= 1_2 \otimes \sigma_2.
\end{align}
Then, the chiral operator on $\mathbf{S}^{\prime \prime}$ is 
\begin{align}
    \bar{\gamma}^{\prime \prime}= (-i \gamma^3 \gamma^4)|_{\mathbf{S}^{\prime \prime}}= \sigma_3 ,
\end{align}
and the transformation acts on $\mathbf{S}^{\prime \prime}$ as
\begin{align}
    \rho(h_2, \vec{z}^f )= \exp( \frac{\pi}{2}i \sigma_3)e^{-i\frac{\pi}{2}}\exp( i \pi \mqty(a & 1) \mqty( n_{22} & -2 \alpha_2 \\ -2 \beta_2 & 0 ) \mqty(b \\ 1)). 
\end{align}
Finally, we have the \Lef numbers
\begin{align}
    L(1)= &n_{11} n_{22},\\
    L(h)=&  \int_{M^{h_2}} \frac{\text{tr}_{ \mathbf{S}^{\prime \prime} }( \bar{\gamma}^{\prime \prime} \rho(h_2, \vec{z}^f ) ) }{\text{det}_N (1-h_2 )}e^{i \frac{F}{2 \pi}}= n_{11} \frac{1}{2} \sum_{a,b=0,1} \exp( i \pi \mqty(a & 1) \mqty( n_{22} & -2 \alpha_2 \\ -2 \beta_2 & 0 ) \mqty(b \\ 1)),
\end{align}
and the orbifold indices
\begin{align}
    \text{ind}_{T^4/\mathbb{Z}_2}(h_2=1)&= n_{11} \times \text{ind}_{T^2/\mathbb{Z}_2}(h_2 |_\mathcal{N}=1), \\
    \text{ind}_{T^4/\mathbb{Z}_2}(h_2=-1)&=n_{11}   \times \text{ind}_{T^2/\mathbb{Z}_2}(h_2 |_\mathcal{N}=-1).
\end{align}

\subsubsection{$(T^4/\mathbb{Z}_2)_3$}

The third type of $T^4/\mathbb{Z}_2$ is given by $h_3=\mqty(\sigma_1 & 0 \\ 0 &\sigma_1)$. The complex structure $\Omega$ is invariant under the transformation $h_3$ when
\begin{align}
    \sigma_1 \Omega \sigma_1= \Omega.
\end{align}
Here, we suppose that the form of $\Omega$ is
\begin{align}
    \Omega= \mqty( \tau_1 & \tau_2  \\ \tau_2   & \tau_1)
\end{align}
with $\Im (\tau_1 \pm \tau_2)>0  $. $h_3$ acts on the Dirac spinor $\psi$ as
\begin{align}
    h_3 \psi(\vec{z},\Omega)=\exp( \frac{\pi}{2} \frac{\gamma^1 - \gamma^3}{\sqrt{2}}\frac{\gamma^2 - \gamma^4}{\sqrt{2}})\exp( -i\frac{\pi}{2} )\psi( \sigma_1 \vec{z},\Omega).
\end{align}

From the transformation laws of the SS phases given in \eqref{eq:action on SS1} and \eqref{eq:action on SS2}, we impose $\alpha_1= \alpha_2 $ and $\beta_1=\beta_2$. In contrast to the other examples, the SS phases are not quantized. The flux must satisfy
\begin{align}
    0=\mqty( -p^x\epsilon & {}^tN \\-N & p^y \epsilon)  - {}^t \bar{h}^{-1}_2\mqty( -p^x \epsilon& {}^tN \\-N &- p^y\epsilon ) \bar{h}^{-1}_2=\left(\begin{matrix}0 & 2 p^{x} & n_{11} - n_{22} & - n_{12} + n_{21}\\- 2 p^{x} & 0 & n_{12} - n_{21} & - n_{11} + n_{22}\\- n_{11} + n_{22} & - n_{12} + n_{21} & 0 & 2 p^{y}\\n_{12} - n_{21} & n_{11} - n_{22} & - 2 p^{y} & 0\end{matrix}\right)
\end{align}
so the form is given by
\begin{align}
    F= 2 \pi(m (d{x^1} \wedge d{y^1}+d{x^2} \wedge d{y^2})+ n (d{x^1} \wedge d{y^2}+d{x^2} \wedge d{y^1})),
\end{align}
where $m $ and $n$ are integers.

We find that fixed points
\begin{align}
    z^f= x\mqty( 1 \\ 1 )+ \Omega y \mqty(1 \\ 1)= (x+ (\tau_1+\tau_2)y )\mqty(1 \\ 1)
\end{align}
form a single torus. Here, $x$ and $y$ are coordinates on this torus, with identifications $x+1 \sim x$ and $y+1 \sim y$.

As in the previous discussion, we have the \Lef numbers 
\begin{align}
L(1)=m^2-n^2, \\
L(h)= (m+n) ,
\end{align}
and the orbifold indices
\begin{align}
    \text{ind}_{T^4/\mathbb{Z}_2}(h_3=1)&= \frac{1}{2}(m^2 +m -n^2+n ), \\
    \text{ind}_{T^4/\mathbb{Z}_2}(h_3=-1)&=\frac{1}{2}(m^2 -m -n^2-n ).
\end{align}
These results are consistent with Ref.~\cite{Kikuchi:2022psj}.

\subsection{ $T^4/ \mathbb{Z}_3$}

In this subsection, we consider the $T^4/ \mathbb{Z}_3$ generated by a cyclic group of order three. Cube roots of the identity element in $Sp(4,\mathbb{Z})$ are classified into three types:
\begin{align}
    h_1= \mqty(- 1_2 & -1_2 \\ 1_2 & 0),~h_2=\mqty( 0 & \sigma_1 \\ -\sigma_1 & -1_2),~ h_3=\left(\begin{array}{cc|cc}
-1 & 0 & -1 & 0 \\ 
0 & 1 & 0 & 0 \\ \cline{1-4}
1 &0 &0 & 0 \\
0 & 0 & 0 & 1  
\end{array}\right).
\end{align}

\subsubsection{$(T^4/ \mathbb{Z}_3)_1$}

Let $(T^4/ \mathbb{Z}_3)_1$ be an orbifold constructed by $h_1$. The invariant complex structure $\Omega$ is given by
\begin{align}
    \Omega= \frac{1}{2}(-1+ i \sqrt{3})1_2.
\end{align}
The transformation $h_1$ acts on $\psi$ as
\begin{align}
    h_1 \psi(\vec{z},\Omega)= \exp(\frac{\pi}{3} \gamma^1 \gamma^2)e^{-i\frac{\pi}{3}} \exp(\frac{\pi}{3} \gamma^3 \gamma^4)e^{-i\frac{\pi}{3}} \psi( e^{i \frac{2\pi}{3} }\vec{z},\Omega).
\end{align}
%$$ \left(\begin{matrix}0 & - n_{12} + n_{21} - p^{y} & 0 & - n_{12} + n_{21} - p^{x}\\n_{12} - n_{21} + p^{y} & 0 & n_{12} - n_{21} + p^{x} & 0\\0 & - n_{12} + n_{21} - p^{x} & 0 & - p^{x} + p^{y}\\n_{12} - n_{21} + p^{x} & 0 & p^{x} - p^{y} & 0\end{matrix}\right) $$
From the invariant conditions for the flux and SS phases, we have
\begin{align}
    F= 2\pi \qty( (n_{21}- n_{12})(dx^1 \wedge dx^2 +dy^1 \wedge dy^2  ) + \mqty(dx^1 ,dx^2) \wedge \mqty(n_{11} & n_{21} \\ n_{12}  & n_{22})  \mqty(dy^1 \\ dy^2 )),
\end{align}
and 
\begin{align}
    \alpha_1= \beta_1= n_{11}/6 +u/3 ,~\alpha_2= \beta_2= n_{22}/6 +v/3, 
\end{align}
where $u,~ v \in \mathbb{Z}$.

We find the nine fixed points 
\begin{align}
    z^f= \frac{1}{3} \qty( \mqty(c_1 \\c_2) -\Omega \mqty(c_1 \\c_2) ),
\end{align}
where $c_1,~c_2=0,~1,~2$. At these points, $h_1$ leads to
\begin{align}
    \rho(h_1, z^f)=& \exp(\frac{\pi}{3} \gamma^1 \gamma^2)e^{-i\frac{\pi}{3}} \exp(\frac{\pi}{3} \gamma^3 \gamma^4)e^{-i\frac{\pi}{3}} \nonumber \\
    &\times \exp(i\pi \mqty( c_1& c_2& 1) \mqty( -\frac{n_{11}}{3} & - \frac{2 n_{12}}{3} + \frac{ n_{21}}{3} & + \frac{n_{11}}{6} +\frac{u}{3} \\
    - \frac{2 n_{12}}{3} + \frac{ n_{21}}{3} & -\frac{n_{22}}{3} &  + \frac{n_{22}}{6} + \frac{v}{3} \\
     + \frac{n_{11}}{6} + \frac{u}{3} &  +\frac{n_{22}}{6} + \frac{v}{3} & 0 ) \mqty(c_1 \\ c_2 \\1 )).
\end{align}
%$$e^{i \pi \left(- \frac{c_{1}^{2} n_{11}}{3} + c_{1} c_{2} \left(- \frac{4 n_{12}}{3} + \frac{2 n_{21}}{3}\right) + 2 c_{1} \left(\frac{n_{11}}{6} + \frac{u}{3}\right) - \frac{c_{2}^{2} n_{22}}{3} + 2 c_{2} \left(\frac{n_{22}}{6} + \frac{v}{3}\right)\right)}$$

Then, we have the \Lef numbers
\begin{align}
L(1)=&- (n_{21}- n_{12})^2 + n_{11} n_{22}- n_{12}n_{21}, \\
    L(h_1)=&-\frac{\omega^{-1}}{3} \sum_{c_1, c_2=0,1,2} \exp(i\pi \mqty( c_1& c_2& 1) \mqty( -\frac{n_{11}}{3} & - \frac{2 n_{12}}{3} + \frac{ n_{21}}{3} & + \frac{n_{11}}{6} +\frac{u}{3} \\
    - \frac{2 n_{12}}{3} + \frac{ n_{21}}{3} & -\frac{n_{22}}{3} &  + \frac{n_{22}}{6} + \frac{v}{3} \\
     + \frac{n_{11}}{6} + \frac{u}{3} &  +\frac{n_{22}}{6} + \frac{v}{3} & 0 ) \mqty(c_1 \\ c_2 \\1 )), \\
     L(h_1^2)=&-\frac{\omega^{-2}}{3} \sum_{c_1, c_2=0,1,2} \exp(i2\pi \mqty( c_1& c_2& 1) \mqty( -\frac{n_{11}}{3} & - \frac{2 n_{12}}{3} + \frac{ n_{21}}{3} & + \frac{n_{11}}{6} +\frac{u}{3} \\
    - \frac{2 n_{12}}{3} + \frac{ n_{21}}{3} & -\frac{n_{22}}{3} &  + \frac{n_{22}}{6} + \frac{v}{3} \\
     + \frac{n_{11}}{6} + \frac{u}{3} &  +\frac{n_{22}}{6} + \frac{v}{3} & 0 ) \mqty(c_1 \\ c_2 \\1 )),
\end{align}
and the orbifold indices 
\begin{align}
    \text{ind}_{T^4/\mathbb{Z}_3}(h_1= \omega^k)= \frac{1}{3}\qty(\sum_{l=0}^2 L(h_1^l) \omega^{-kl}).
\end{align}
We have confirmed that these values are integers through a brute-force calculation, and they coincide
with the results obtained in the previous work \cite{Kikuchi:2022psj}. 

\subsubsection{$(T^4/ \mathbb{Z}_3)_2$}
The second type of $T^4/ \mathbb{Z}_3$ is constructed by $h_2=\mqty( 0 & \sigma_1 \\ -\sigma_1 & -1_2)$. The invariant complex structure $\Omega$ under $h_2$ is given by
\begin{align}
    \Omega= \frac{1}{2}(- \sigma_1 + i \sqrt{3} )= \mqty(i\frac{\sqrt{3}}{2} & -\frac{1}{2} \\ -\frac{1}{2} & i\frac{\sqrt{3}}{2}),
\end{align}
and the action on the Dirac spinor is 
\begin{align}
    h_2\psi (\vec{z},\Omega)= \exp( -\frac{\pi}{3} \gamma^{\prime1} \gamma^{\prime2}) \exp(+\frac{\pi}{3} \gamma^{\prime3} \gamma^{\prime4})\psi( {}^t(-\sigma_1 \Omega-1_2)\vec{z},\Omega), 
\end{align}
where $\gamma^{\prime}$ are given by
\begin{align}
    \mqty( \gamma^{\prime 1} \\ \gamma^{\prime 2} \\ \gamma^{\prime 3} \\ \gamma^{\prime 4})= \frac{1}{2}\left(\begin{matrix}1 & 1 & 1 & 1\\-1 & -1 & 1 & 1\\1 & -1 & 1 & -1\\-1 & 1 & 1 & -1\end{matrix}\right) \mqty( \gamma^{1} \\ \gamma^{ 3} \\ \gamma^{2} \\ \gamma^{ 4}). %この式はただしい
    \end{align}
%$$\left(\begin{matrix}0 & n_{11} - n_{22} + p^{y} & n_{11} - n_{22} - p^{x} & 0\\- n_{11} + n_{22} - p^{y} & 0 & 0 & - n_{11} + n_{22} + p^{x}\\- n_{11} + n_{22} + p^{x} & 0 & 0 & p^{x} + p^{y}\\0 & n_{11} - n_{22} - p^{x} & - p^{x} - p^{y} & 0\end{matrix}\right)$$
The flux and SS phases are restricted by the invariant conditions. Then, we get
\begin{align}
    F= 2\pi \qty( (n_{11}- n_{22}) (-dx^1 \wedge dx^2 +dy^1 \wedge dy^2 )+ \mqty(dx^1 & dx^2) \wedge \mqty(n_{11} & n_{21} \\ n_{12}  & n_{22})  \mqty(dy^1 \\ dy^2 )),
\end{align}
and 
\begin{align}
    \alpha_1= \beta_2= \frac{n_{21}}{6}+ \frac{u}{3},~\alpha_2= \beta_1= \frac{n_{12}}{6}+ \frac{v}{3}.
\end{align}
There are nine fixed points
\begin{align}
    z^f= \frac{1}{3} \qty( \mqty(c_1 \\c_2 ) - \Omega \mqty(c_2 \\ c_1)), 
\end{align}
where $c_1,~c_2=0,~1,~2$.
At the fixed points, $h_2$ acts as
\begin{align}
    \rho(h_2, z^f)=&  \exp(\frac{\pi}{3} \gamma^{\prime1} \gamma^{\prime2}) \exp(-\frac{\pi}{3} \gamma^{\prime3} \gamma^{\prime4}) \nonumber \\
    &\times \exp(i\pi \mqty( c_1& c_2& 1) \mqty( \frac{n_{21}}{3} &  \frac{ 2n_{22}}{3} - \frac{ n_{11}}{3} & - \frac{n_{21}}{6} -\frac{u}{3} \\
     \frac{2 n_{22}}{3} - \frac{ n_{11}}{3} & \frac{n_{12}}{3} &  - \frac{n_{12}}{6} - \frac{v}{3} \\
     -\frac{n_{21}}{6} - \frac{u}{3} &  -\frac{n_{12}}{6} - \frac{v}{3} & 0 ) \mqty(c_1 \\ c_2 \\1 )).
\end{align}

%$$e^{i \pi \left(\frac{c_{1}^{2} n_{21}}{3} + c_{1} c_{2} \left(- \frac{2 n_{11}}{3} + \frac{4 n_{22}}{3}\right) - 2 c_{1} \left(\frac{n_{21}}{6} + \frac{u}{3}\right) + \frac{c_{2}^{2} n_{12}}{3} - 2 c_{2} \left(\frac{n_{12}}{6} + \frac{v}{3}\right)\right)}$$

We use $\gamma^{\prime 1} \cdots\gamma^{\prime 4}= \gamma^{ 1} \cdots\gamma^{ 4}$, and the \Lef numbers is given by
\begin{align}
L(1)=& (n_{11}- n_{22})^2 + n_{11} n_{22}- n_{12}n_{21}, \\
    L(h_2)=&\frac{1}{3} \sum_{c_1, c_2=0,1,2}  \exp(i\pi \mqty( c_1& c_2& 1) \mqty( \frac{n_{21}}{3} &  \frac{ 2n_{22}}{3} - \frac{ n_{11}}{3} & - \frac{n_{21}}{6} -\frac{u}{3} \\
     \frac{2 n_{22}}{3} - \frac{ n_{11}}{3} & \frac{n_{12}}{3} &  - \frac{n_{12}}{6} - \frac{v}{3} \\
     -\frac{n_{21}}{6} - \frac{u}{3} &  -\frac{n_{12}}{6} - \frac{v}{3} & 0 ) \mqty(c_1 \\ c_2 \\1 )),\\
     L(h_2^2)=&\frac{1}{3} \sum_{c_1, c_2=0,1,2} \exp(i2\pi \mqty( c_1& c_2& 1) \mqty( \frac{n_{21}}{3} &  \frac{ 2n_{22}}{3} - \frac{ n_{11}}{3} & - \frac{n_{21}}{6} -\frac{u}{3} \\
     \frac{2 n_{22}}{3} - \frac{ n_{11}}{3} & \frac{n_{12}}{3} &  - \frac{n_{12}}{6} - \frac{v}{3} \\
     -\frac{n_{21}}{6} - \frac{u}{3} &  -\frac{n_{12}}{6} - \frac{v}{3} & 0 ) \mqty(c_1 \\ c_2 \\1 )).
\end{align}
We have confirmed that these values are integers through a brute-force calculation, and they coincide
with the results obtained in the previous work \cite{Kikuchi:2022psj}.

\begin{comment}

\begin{align}
    g= \mqty( 0 & -\sigma_3 \\ \sigma_3 & -I)
\end{align}

\begin{align}
    F= 2\pi \qty( p(dx^1 \wedge dx^2 -dy^1 \wedge dy^2  ) + \mqty(dx^1 ,dx^2) \wedge \mqty(n & m- p/2 \\ -m-p/2  & l)  \mqty(dy^1 \\ dy^2 ))
\end{align}

%小林さんのパラメータを$n_k,~m_k$と書くと、$p=-2m_k,~ n=n_k+ 2m_k,~m=0,~l=n_k -2m_k$と対応する。%m_kの符号は確認するべき

SS phase
\begin{align}
    \alpha_1+ \beta_1 \in \mathbb{Z},~\alpha_2- \beta_2 \in \mathbb{Z}
\end{align}

\begin{align}
    \beta_1=-n/6+u/3,~\beta_2=l/6+v/3
\end{align}

固定点
\begin{align}
    \vec{x}= \vec{y}= \frac{1}{3} \mqty(b_1 \\  1-b_2 )~(b_1,~b_2 =0,~1,~2)
\end{align}

\Lef 数
\begin{align}
    L(1)= p^2+nl+(m-p/2)(m+p/2)
\end{align}

\begin{align}
    L(g)= +\frac{1}{3} \sum_{b_1,b_2=0,1,2} \exp( i \pi (b_1,b_2,1) 
    \mqty( n/3 & m/3 +p/2 & n/6 -u/3 \\
    m/3+p/2 & -l/3 & l/6+v/3 \\
    n/6 -u/3 & l/6 +v/3 & 0
     ) \mqty(b_1 \\ b_2 \\1))
\end{align}
ここの符号は確認

\begin{align}
    \text{ind}(1)&=\frac{1}{3}\mqty( L(1) + L(g) + L(g^2) ) =D(+1)\\
    \text{ind}(w)&=\frac{1}{3}\mqty( L(1) + w^2 L(g) + w L(g^2) ) =D(\omega) \\
    \text{ind}(w^2)&=\frac{1}{3}\mqty( L(1) + w L(g) + w^2 L(g^2) ) =D(\omega^2) 
\end{align}

小林さんのパラメータを$n_k,~m_k$と書くと、$p=-2m_k,~ n=n_k+ 2m_k,~m=0,~l=n_k -2m_k,~\beta_1=n_k/2,~\beta_2=n_k/2$と対応する。\fix{あとで確認}
\end{comment}
\subsubsection{$(T^4/ \mathbb{Z}_3)_3$}
We consider the third type of $T^4/ \mathbb{Z}_3$ constructed by
\begin{align}
h_3=\left(\begin{array}{cc|cc}
-1 & 0 & -1 & 0 \\ 
0 & 1 & 0 & 0 \\ \cline{1-4}
1 &0 &0 & 0 \\
0 & 0 & 0 & 1  
\end{array}\right).
\end{align}
The invariant complex structure $\Omega$ is given by
\begin{align}
    \Omega= \mqty( \frac{1}{2}(-1 +i \sqrt{3}) & 0 \\ 0 & \tau),
\end{align}
where $\tau$ is a complex number such that $\Im \tau>0$. $h_3$ acts on the Dirac spinor as
\begin{align}
    h_3 \psi(\vec{z},\Omega)= \exp(\frac{\pi}{3} \gamma^1 \gamma^2) e^{-i\frac{\pi}{3}} \psi( \mqty(\omega & 0\\ 0& 1)\vec{z},\Omega ),
\end{align}
and the flux is restricted as
\begin{align}
    F= 2\pi(n_{11} dx^1 \wedge dy^1 + n_{22} dx^2 \wedge dy^2 ).
\end{align}
%$$\left(\begin{matrix}0 & n_{12} + p^{x} & 0 & n_{21} - p^{y}\\- n_{12} - p^{x} & 0 & 2 n_{12} - p^{x} & 0\\0 & - 2 n_{12} + p^{x} & 0 & n_{21} + 2 p^{y}\\- n_{21} + p^{y} & 0 & - n_{21} - 2 p^{y} & 0\end{matrix}\right)$$
The orbifold indices are obtained as
\begin{align}
    \text{ind}_{T^4/\mathbb{Z}_3}(h_3= \omega^k)=\text{ind}_{T^2/\mathbb{Z}_3 }(h_3 |_{M^{h_3} }=\omega^k )n_{22}.
\end{align}

\subsection{$T^4/ \mathbb{Z}_5$}
In this subsection, we consider the $T^4/ \mathbb{Z}_5$ generated by a cyclic group of order five. The generator is given by
\begin{align}
    h=\left(\begin{array}{cc|cc}
-1 & -1 & -1 & 0 \\ 
-1 & 0 & 0 & -1 \\ \cline{1-4}
1 &0 &0 & 0 \\
0 & 1 & 0 & 0  
\end{array}\right).
\end{align}
%これでhが尽きていることの説明

%$$\left(\begin{matrix}- \frac{1}{2} + \frac{i \sqrt{\frac{2 \sqrt{5}}{5} + 1}}{2} & - \frac{1}{2} - \frac{i \sqrt{1 - \frac{2 \sqrt{5}}{5}}}{2}\\- \frac{1}{2} - \frac{i \sqrt{1 - \frac{2 \sqrt{5}}{5}}}{2} & \frac{i \sqrt{\frac{2 \sqrt{5}}{5} + 2}}{2}\end{matrix}\right)$$
We have the invariant complex structure $\Omega$ under $h$ as
\begin{align}
    \Omega= \frac{1}{2} \mqty(- 1+  i\sqrt{ \frac{5 + 2 \sqrt{5}}{5}} &-1 -i \sqrt{ \frac{5 - 2 \sqrt{5}}{5}} \\-1 -i\sqrt{ \frac{5 - 2 \sqrt{5}}{5}} & i\sqrt{ \frac{10 + 2 \sqrt{5}}{5}} ) ,
\end{align}
and the action on the Dirac spinor as
\begin{align}
    h\psi(\vec{z},\Omega)= \exp(\frac{\pi}{5} \frac{\gamma^1 -\phi \gamma^3}{\sqrt{2+ \phi} } \frac{\gamma^2 -\phi \gamma^4}{\sqrt{2+ \phi} }  ) \exp(2\frac{\pi}{5} \frac{\phi \gamma^1 + \gamma^3}{\sqrt{2+ \phi} } \frac{\phi \gamma^2 +\gamma^4}{\sqrt{2+ \phi} }  ) 
e^{-i \frac{3\pi}{5}} \psi( \Omega\vec{z},\Omega),
\end{align}
where $\phi=\frac{1}{2}(1+ \sqrt{5})$ is the golden ratio.
%$$\left(\begin{matrix}0 & p^{x} - p^{y} & p^{y} & - n_{12} + n_{21}\\- p^{x} + p^{y} & 0 & n_{12} - n_{21} - p^{y} & - p^{y}\\- p^{y} & - n_{12} + n_{21} + p^{y} & 0 & - n_{11} + n_{12} + n_{22} - p^{x} + 2 p^{y}\\n_{12} - n_{21} & p^{y} & n_{11} - n_{12} - n_{22} + p^{x} - 2 p^{y} & 0\end{matrix}\right)$$
For consistency, the flux must be in the following form
\begin{align}
    F= 2\pi  \mqty(dx^1 & dx^2) \wedge \mqty( n_{22}+ n_{12} & n_{12} \\ n_{12} & n_{22} ) \mqty(dy^1 \\ dy^2).
\end{align}
The SS phases must satisfy
\begin{align}
    \alpha_1=\beta_1= \frac{1}{10}(3n_{12}- 2 n_{22})-\frac{2u}{5},~\alpha_2=\beta_2= \frac{1}{10}(n_{12}+n_{22}) +\frac{u}{5}\quad (u \in \mathbb{Z}).
\end{align}
Under this transformation, we have the following fixed points
\begin{align}
    \vec{z}^f= \frac{1}{5} \qty( \mqty(c \\ 2c) -\Omega \mqty(c \\ 2c))\quad (c=0,1,2,3,4).
\end{align}
At the fixed points, $h$ acts as
\begin{align}
    \rho(h^k,\vec{z}^f)=\exp(k\frac{\pi}{5} \frac{\gamma^1 -\phi \gamma^3}{\sqrt{2+ \phi} } \frac{\gamma^2 -\phi \gamma^4}{\sqrt{2+ \phi} }  ) \exp(2k\frac{\pi}{5} \frac{\phi \gamma^1 + \gamma^3}{\sqrt{2+ \phi} } \frac{\phi \gamma^2 +\gamma^4}{\sqrt{2+ \phi} }  ) 
e^{-i k\frac{3\pi}{5}},
\end{align}
so the Lefschetz numbers are obtained as
\begin{align}
    L(1)= &n_{22}(n_{22}+n_{12})-n_{12}^2, \\
    L(h^k)=&-\frac{4}{5} \sum_{c=0}^4\sin( k\frac{\pi }{5}) \sin( k\frac{2\pi }{5})   \exp(ik\frac{\pi }{5}( c^2(4n_{12} +3n_{22}) +c (4n_{12} -n_{22}-2u) -3) ).
\end{align}
We have confirmed that these values are integers through a brute-force calculation, and they coincide
with the results obtained in the previous work \cite{Kikuchi:2022psj}. 

\section{Coxeter Orbifold Model}
\label{sec:Coxeter}

In this section, we consider a four-dimensional orbifold generated by an element that is not generally expressed by modular transformations. We assume that the lattice $\Lambda$ is a root lattice and construct a four-dimension torus as $\mathbb{R}^4/\Lambda$. Let $s_1,\cdots,s_4 \in \mathbb{R}^4$ span the root lattice $\Lambda$. We then define a reflection with respect to the hyperplane perpendicular to $s_i$ by
\begin{align}
    \text{Ref}_{s_i}(v) = v - 2\frac{(v, s_i)}{(s_i, s_i)}s_i.
\end{align}
Note that all the reflections preserve the Euclidean metric on $\mathbb{R}^4$ and map $\Lambda$ to itself. The Coxeter element $C$ is obtained as the product of all the reflections with respect to $s_i$:
\begin{align}
    C= \text{Ref}_{s_1} \cdots \text{Ref}_{s_4}.
\end{align}
This definition depends on the choice of basis, but all Coxeter elements are conjugate to each other. Mathematically, there exists an integer number $h$ such that $C^h=1_4$, and $C$ generates a cyclic group of order $h$. Then, a Coxeter orbifold is defined by $C$. As we will see below, $C$ cannot be described by modular transformations. 

Here, we consider the $D_4$ lattice, which is generated by
\begin{align}
    e_1=\mqty(1 \\ 0 \\ 0 \\ 0),~
    f_1=\mqty(0 \\ 1 \\ 0 \\ 0)=e_3,~
    e_2=\mqty(-\frac{1}{2} \\ -\frac{1}{2} \\ \frac{1}{\sqrt{2}}   \\ 0),~
    f_2=\mqty(-\frac{1}{2} \\  -\frac{1}{2}  \\ 0 \\ \frac{1}{\sqrt{2}} )=e_4.
\end{align}
According to \cite{Kikuchi:2022psj}, this lattice is associated with a complex structure 
\begin{align}
    \Omega= \mqty( -\frac{1}{2} +i \frac{1}{\sqrt{2}} &  -\frac{1}{2}  \\ -\frac{1}{2}  & -\frac{1}{2} +i \frac{1}{\sqrt{2}} ),
\end{align}
which appears in $T^4/\mathbb{Z}_8$ modular orbifold model. Note that the Dynkin diagram of $D_4$ is represented by $e_1,~ e_2,~e_3$, and $-(e_1+e_2+e_3+e_4)$. We assign the real coordinates of $\mathbb{R}^4/\Lambda$ as
\begin{align}
    z=x^1e_1 +y^1 e_2+ x^2 e_3 +y^2e_4.
\end{align}
Then, the Coxeter element of $D_4$ is expressed as
\begin{align}
C&= \text{Ref}_{e_1}  \text{Ref}_{e_3} \text{Ref}_{e_2} \text{Ref}_{-(e_1+e_2+e_3+e_4)} ,\\
    C(z)&=\mqty(e_1 & e_3 & e_2 & e_4)\left(\begin{array}{cc|cc}
        0 & 1 & -1 & 0 \\ 
        1 & 0 & -1 & 0 \\ \cline{1-4}
        1 & 1 & -1 & -1 \\
        0 & 0 & -1& 0  
        \end{array}\right) \mqty(x^1 \\ x^2 \\ y^1 \\ y^2).
\end{align}
Since
\begin{align}
    {}^t C \mqty( 0  & -1_n \\ 1_n & 0) C \neq   \mqty( 0  & -1_n \\ 1_n & 0),
\end{align}
the Coxeter element does not belong to the symplectic group $Sp(4,\mathbb{Z})$. The Coxeter element acts on the Dirac spinor as
\begin{align}
    C\psi(z)= \rho(C) \psi(C^{-1}z),
\end{align}
where $\rho(C)$ is a spin representation defined by
\begin{align}
    \rho(C)= \gamma^1 \gamma^3 \qty(-\frac{1}{2}\gamma^1 -\frac{1}{2}\gamma^3+ \frac{1}{\sqrt{2}}\gamma^2 )  \qty(-\frac{1}{\sqrt{2}}\gamma^2-\frac{1}{\sqrt{2}}\gamma^4).
\end{align}
There is an ambiguity in choosing the $U(1)$ phase factor. As in the previous discussions, the invariant flux is written as
\begin{align}
    F=2\pi \qty( p^x dx^1 \wedge dx^2 + p^y dy^1 \wedge dy^2 + \mqty(dx^1 & dx^2) \wedge \mqty( -p^x+p^y & -p^x-p^y \\ p^x +p^y & p^x -p^y) \mqty(dy^1 \\ dy^2)  ).
\end{align}

The sixth power of $C$ is the identity element, and we find three cyclic groups: $\mathbb{Z}_2,~\mathbb{Z}_3$, and $\mathbb{Z}_6$, generated by $C^3,~C^2$, and $C$, respectively. However, since $C^3=-1_4$, the orbifold is the same as in Section \ref{subsec:T4Z2_1}. We denote the orbifolds constructed by $C^2$ and $C$ as $(T^4/\mathbb{Z}_3)_{D_4}$ and $(T^4/\mathbb{Z}_6)_{D_4}$, respectively.  %In the remaining section, we calculate the orbifold indices in $T^4/ \mathbb{Z}_3$ a 

%For example, if the lattice $\Lambda$ is a root lattice, we find a Coxe

%Such an orbifold is called Coxeter orbifold. In the Coxeter orbifold, the torus $T^4$ is given by $\mathbb{R}^4/ \Lambda$, where $\Lambda$ is a root lattice. 

%$$\left(\begin{matrix}0 & n_{21} - n_{22} + 2 p^{x} & n_{11} + n_{22} & n_{12} - p^{x} - p^{y}\\- n_{21} + n_{22} - 2 p^{x} & 0 & n_{12} + n_{21} & n_{11} + p^{x} - p^{y}\\- n_{11} - n_{22} & - n_{12} - n_{21} & 0 & n_{21} + n_{22} + 2 p^{y}\\- n_{12} + p^{x} + p^{y} & - n_{11} - p^{x} + p^{y} & - n_{21} - n_{22} - 2 p^{y} & 0\end{matrix}\right)$$

\subsection{$(T^4/\mathbb{Z}_3)_{D_4}$}

In this subsection, we calculate the orbifold indices on $(T^4/\mathbb{Z}_3)_{D_4}$. The generator is given by
\begin{align}
h=C^2=\left(\begin{array}{cc|cc}
0 & -1 & 0 & 1 \\ 
-1 & 0 & 0 & 1 \\ \cline{1-4}
0 & 0 &0 & 1 \\
-1 & -1 & 1&   1
\end{array}\right).
\end{align}
This operator acts on the Dirac spinor as
\begin{align}
    h\psi(z,\Omega)= \rho(C^2) \psi(C^{-2} z, \Omega),
\end{align}
and the consistent SS phases are given by
\begin{align}
    \alpha_2= -\alpha_1-\beta_1-\frac{p^x+p^y}{2},~\beta_2=\beta_1.
\end{align}

We find the fixed points as
\begin{align}
    \vec{z}_{\mathbb{Z}_3}^f= x\mqty(1 \\ -1 \\ 0 \\ 0 )+y \mqty( 1 \\ 0 \\ 1 \\ 1 ). \label{eq:Coxeter_Z3_fixed_point}
\end{align}
Since the parameters $x$ and $y$ are identified by $x+1$ and $y+1$, the set of the fixed points form a single torus $T^2$. By using the Atiyah-Segal-Singer fixed point theorem, we have the \Lef numbers
\begin{align}
    L(1)=3p^x p^y ,~
    L(h)=-i\sqrt{3}p^x ,~
    L(h^2)=i\sqrt{3}p^x,
\end{align}
and the orbifold indices
\begin{align}
    \text{ind}_{T^4/\mathbb{Z}_3} (h=1)= p^xp^y,~
    \text{ind}_{T^4/\mathbb{Z}_3} (h=\omega)= p^x (p^y-1),~
    \text{ind}_{T^4/\mathbb{Z}_3} (h=\omega^2)= p^x( p^y+1).
\end{align}

\subsection{$(T^4/\mathbb{Z}_6)_{D_4}$}

In this subsection, we consider $(T^4/\mathbb{Z}_6)_{D_4}$ generated by
\begin{align}
    h=C=\left(\begin{array}{cc|cc}
        0 & 1 & -1 & 0 \\ 
        1 & 0 & -1 & 0 \\ \cline{1-4}
        1 & 1 & -1 & -1 \\
        0 & 0 & -1& 0  
        \end{array}\right).
\end{align}
The consistent SS phases are given by
\begin{align}
\alpha_1= \frac{u}{2},~\beta_1=\frac{v}{2},~
    \alpha_2= -\alpha_1-\beta_1-\frac{p^x+p^y}{2},~\beta_2=\beta_1,
\end{align}
where $u,~v\in \mathbb{Z}$.

%$$\left(\begin{matrix}0 & n_{21} - n_{22} + 2 p^{x} & n_{11} + n_{22} & n_{12} - p^{x} - p^{y}\\- n_{21} + n_{22} - 2 p^{x} & 0 & n_{12} + n_{21} & n_{11} + p^{x} - p^{y}\\- n_{11} - n_{22} & - n_{12} - n_{21} & 0 & n_{21} + n_{22} + 2 p^{y}\\- n_{12} + p^{x} + p^{y} & - n_{11} - p^{x} + p^{y} & - n_{21} - n_{22} - 2 p^{y} & 0\end{matrix}\right)$$

We find four $\mathbb{Z}_6$-fixed points 
\begin{align}
    \vec{z}^f_{\mathbb{Z}_6}= \frac{a}{2}\mqty(1 \\ -1 \\ 0 \\ 0 )+\frac{b}{2}\mqty( 1 \\ 0 \\ 1 \\ 1 )\quad (a,~b\in \mathbb{Z}).
\end{align}
On these fixed points, $h$ acts as
\begin{align}
    \rho(h,z^f_{\mathbb{Z}_6})= \rho(C) \exp(i\pi \mqty(a& b&1) \mqty( -p^x & -\frac{p^x}{2} & -\frac{p^x+p^y+u}{2} \\  -\frac{p^x}{2}& -\frac{p^y}{2}   & -\frac{v}{2} \\  -\frac{p^x+p^y+u}{2}  & -\frac{v}{2} &0 ) \mqty(a\\ b\\1)).
\end{align}
We also discover $\mathbb{Z}_3$-fixed points as shown in \eqref{eq:Coxeter_Z3_fixed_point} and $16$ $\mathbb{Z}_2$-fixed points as
\begin{align}
    z^f_{\mathbb{Z}_2}=\frac{1}{2} \mqty(c_1\\ c_2 \\ c_3 \\ c_4),
\end{align}
where $c_i=0,~1$. Then, we have the \Lef numbers
\begin{align}
    L(1)&=3p^x p^y ,\\ 
    L(h)&=-\frac{1}{2}\exp(i\pi \mqty(a& b&1) \mqty( -p^x & -\frac{p^x}{2} & -\frac{p^x+p^y+u}{2} \\  -\frac{p^x}{2}& -\frac{p^y}{2}   & -\frac{v}{2} \\  -\frac{p^x+p^y+u}{2}  & -\frac{v}{2} &0 ) \mqty(a\\ b\\1)), \\
    L(h^2)&= -i\sqrt{3}p^x,  \\
    L(h^3)&= \frac{1}{4}\sum_{c_1,c_2,c_3,c_4=0,1} \exp( i\pi \qty( p^x c_1 c_2+p^y c_3 c_4 + \mqty(c_1 &c_2 & 1) \mqty( -p^x+ p^y & -p^x-p^y & -2 \alpha_1  \\
    p^x+ p^y & p^x-p^y & -2 \alpha_2   \\
    -2 \beta_1 & -2 \beta_2 & 0) \mqty(c_3 \\ c_4 \\ 1) )), \\
    L(h^4)&=i\sqrt{3}p^x,  \\
    L(h^5)&=-\frac{1}{2}\exp(-i\pi \mqty(a& b&1) \mqty( -p^x & -\frac{p^x}{2} & -\frac{p^x+p^y+u}{2} \\  -\frac{p^x}{2}& -\frac{p^y}{2}   & -\frac{v}{2} \\  -\frac{p^x+p^y+u}{2}  & -\frac{v}{2} &0 ) \mqty(a\\ b\\1)) ,
\end{align}
and the orbifold indices using Eq. \eqref{eq:formula_of_index}.

%$$e^{i \pi \left(c_{1} c_{2} p^{x} - c_{1} d_{1} p^{x} + c_{1} d_{1} p^{y} - c_{1} d_{2} p^{x} - c_{1} d_{2} p^{y} - c_{1} v + c_{2} d_{1} p^{x} + c_{2} d_{1} p^{y} + c_{2} d_{2} p^{x} - c_{2} d_{2} p^{y} - 2 c_{2} \left(- \frac{p^{x}}{2} - \frac{p^{y}}{2} - \frac{u}{2} - \frac{v}{2}\right) + d_{1} d_{2} p^{y} - d_{1} u + d_{2} u\right)}$$

\section{Summary}
\label{Conclusion}
We have demonstrated that the index on an orbifold can be calculated using the fixed point theorem. The necessary information for the calculation includes the fixed points of the orbifold and how the spatial symmetries act on these points, eliminating the need to compute zero modes as required in previous studies. Since the fixed point theorem can be applied to any fermion theory on any orbifold, it allows for the determination of the index even on orbifolds where the calculation of zero modes is difficult or in non-trivial gauge configurations. This method is an extension of the zero mode counting method discussed in previous studies \cite{Sakamoto:2020pev,Kobayashi:2022tti,Imai:2022bke,Kikuchi:2022lfv,Kikuchi:2022psj}.

In Section \ref{Section2}, we have reviewed the fixed point theorem and the Lefschetz number.
In Section \ref{sec:Setup}, we have constructed the magnetized 
$T^{2n}$ model, determined the metric to be compatible with the modular transformation group, and derived the spin$^c$ representation, 
$U(1)$ gauge field, and SS phase transformation rules so that they commute with the Dirac operator.
In Section \ref{T2ZN}, as a concrete example, we have derived the index of the magnetized 
$T^{2}/\mathbb{Z}_N\,(N=2,3,4,6)$ orbifold from the fixed point theorem and confirmed that it matches with previous studies \cite{Sakamoto:2020pev,Kobayashi:2022tti,Imai:2022bke}.
In Section \ref{T4ZN}, we have derived the index of the magnetized 
$T^{4}/\mathbb{Z}_N$
  orbifold from the fixed point theorem and confirmed that it matches with previous studies \cite{Kikuchi:2022lfv,Kikuchi:2022psj}. Although there are countless cyclic subgroups to create orbifolds, the equivalence classes under similarity transformation are at most finite, and it is sufficient to consider the simplest generator within the equivalence class. Furthermore, we can calculate the index in parameter regions not considered in previous studies \cite{Kikuchi:2022lfv,Kikuchi:2022psj}.
In Section \ref{sec:Coxeter}, we also calculated the index of the orbifold generated by the Coxeter element which cannot be described by modular transformation. Physically interesting orbifolds are created by Coxeter elements associated with root lattices, and have been studied for a long time in mathematics \cite{Dixon:1985jw, Dixon:1986jc, Bailin:1999nk}. Using our method, it becomes possible to calculate the orbifold index for all root lattices. As far as we know, the calculation of the orbifold index generated by the Coxeter element, which cannot be described by modular transformation, is a new result.

This study has made it possible to determine the number of chiral zero modes in fermion theories on any orbifold, thereby elucidating the generation structure in compactifications with singularities. This is a powerful tool for classifying compactifications that can yield realistic chiral theories. Additionally, it is significant that the physical interpretation of the Lefschetz number in mathematics has been clarified.

Phenomenological applications can also be considered. As discussed in papers \cite{Takeuchi:2023egl,Ishiguro:2020rot}, %\cite{Baur:2024qzo}\fix{?}, 
if the formula for the number of generations is clear, we can discuss the conditions that provide its upper limit. Previously, this was done concerning smooth manifolds, but with this study, it becomes possible to analyze cases with singularities.

\section*{Acknowledgement}
We would like to express our gratitude to Mikio Furuta, Shinichiroh Matsuo, and Makoto Sakamoto. In particular, we are especially thankful to Mikio Furuta for his guidance on $K$-theory. S.A. is supported by JSPS KAKENHI Grant Number JP23KJ1459.

\bibliographystyle{unsrt}
\bibliography{ref}

\begin{thebibliography}{10}

\bibitem{Sakamoto:2020pev}
Makoto Sakamoto, Maki Takeuchi, and Yoshiyuki Tatsuta.
\newblock {Zero-mode counting formula and zeros in orbifold compactifications}.
\newblock {\em Phys. Rev. D}, 102(2):025008, 2020.

\bibitem{Abe:2008sx}
Hiroyuki Abe, Kang-Sin Choi, Tatsuo Kobayashi, and Hiroshi Ohki.
\newblock {Three generation magnetized orbifold models}.
\newblock {\em Nucl. Phys. B}, 814:265--292, 2009.

\bibitem{Abe:2015yva}
Tomo-hiro Abe, Yukihiro Fujimoto, Tatsuo Kobayashi, Takashi Miura, Kenji Nishiwaki, Makoto Sakamoto, and Yoshiyuki Tatsuta.
\newblock {Classification of three-generation models on magnetized orbifolds}.
\newblock {\em Nucl. Phys. B}, 894:374--406, 2015.

\bibitem{Libanov:2000uf}
M.~V. Libanov and Sergey~V. Troitsky.
\newblock {Three fermionic generations on a topological defect in extra dimensions}.
\newblock {\em Nucl. Phys. B}, 599:319--333, 2001.

\bibitem{Frere:2000dc}
J.~M. Frere, M.~V. Libanov, and Sergey~V. Troitsky.
\newblock {Three generations on a local vortex in extra dimensions}.
\newblock {\em Phys. Lett. B}, 512:169--173, 2001.

\bibitem{PhysRevD.65.044004}
Andrey Neronov.
\newblock Fermion masses and quantum numbers from extra dimensions.
\newblock {\em Phys. Rev. D}, 65:044004, Jan 2002.

\bibitem{PhysRevD.73.085007}
Silvestre Aguilar and Douglas Singleton.
\newblock Fermion generations, masses, and mixings in a 6d brane model.
\newblock {\em Phys. Rev. D}, 73:085007, Apr 2006.

\bibitem{Gogberashvili:2007gg}
Merab Gogberashvili, Pavle Midodashvili, and Douglas Singleton.
\newblock {Fermion Generations from 'Apple-Shaped' Extra Dimensions}.
\newblock {\em JHEP}, 08:033, 2007.

\bibitem{Guo:2008ia}
Zhi-qiang Guo and Bo-Qiang Ma.
\newblock {Fermion Families from Two Layer Warped Extra Dimensions}.
\newblock {\em JHEP}, 08:065, 2008.

\bibitem{PhysRevLett.108.181807}
David~B. Kaplan and Sichun Sun.
\newblock Spacetime as a topological insulator: Mechanism for the origin of the fermion generations.
\newblock {\em Phys. Rev. Lett.}, 108:181807, May 2012.

\bibitem{Cremades:2004wa}
D.~Cremades, L.~E. Ibanez, and F.~Marchesano.
\newblock {Computing Yukawa couplings from magnetized extra dimensions}.
\newblock {\em JHEP}, 05:079, 2004.

\bibitem{Arkani-Hamed:1999ylh}
Nima Arkani-Hamed and Martin Schmaltz.
\newblock {Hierarchies without symmetries from extra dimensions}.
\newblock {\em Phys. Rev. D}, 61:033005, 2000.

\bibitem{Dvali:2000ha}
G.~R. Dvali and Mikhail~A. Shifman.
\newblock {Families as neighbors in extra dimension}.
\newblock {\em Phys. Lett. B}, 475:295--302, 2000.

\bibitem{Gherghetta:2000qt}
Tony Gherghetta and Alex Pomarol.
\newblock {Bulk fields and supersymmetry in a slice of AdS}.
\newblock {\em Nucl. Phys. B}, 586:141--162, 2000.

\bibitem{Kaplan:2000av}
David~Elazzar Kaplan and Timothy M.~P. Tait.
\newblock {Supersymmetry breaking, fermion masses and a small extra dimension}.
\newblock {\em JHEP}, 06:020, 2000.

\bibitem{Huber:2000ie}
Stephan~J. Huber and Qaisar Shafi.
\newblock {Fermion masses, mixings and proton decay in a Randall-Sundrum model}.
\newblock {\em Phys. Lett. B}, 498:256--262, 2001.

\bibitem{Kaplan:2001ga}
David~Elazzar Kaplan and Timothy M.~P. Tait.
\newblock {New tools for fermion masses from extra dimensions}.
\newblock {\em JHEP}, 11:051, 2001.

\bibitem{Fujimoto:2012wv}
Yukihiro Fujimoto, Tomoaki Nagasawa, Kenji Nishiwaki, and Makoto Sakamoto.
\newblock {Quark mass hierarchy and mixing via geometry of extra dimension with point interactions}.
\newblock {\em PTEP}, 2013:023B07, 2013.

\bibitem{PhysRevD.97.115039}
Yukihiro Fujimoto, Takashi Miura, Kenji Nishiwaki, and Makoto Sakamoto.
\newblock Dynamical generation of fermion mass hierarchy in an extra dimension.
\newblock {\em Phys. Rev. D}, 97:115039, Jun 2018.

\bibitem{PhysRevD.90.105006}
Hiroyuki Abe, Tatsuo Kobayashi, Keigo Sumita, and Yoshiyuki Tatsuta.
\newblock Gaussian froggatt-nielsen mechanism on magnetized orbifolds.
\newblock {\em Phys. Rev. D}, 90:105006, Nov 2014.

\bibitem{PhysRevD.88.115007}
Yukihiro Fujimoto, Kenji Nishiwaki, and Makoto Sakamoto.
\newblock $cp$ phase from twisted higgs vacuum expectation value in extra dimension.
\newblock {\em Phys. Rev. D}, 88:115007, Dec 2013.

\bibitem{Kobayashi:2016qag}
Tatsuo Kobayashi, Kenji Nishiwaki, and Yoshiyuki Tatsuta.
\newblock {CP-violating phase on magnetized toroidal orbifolds}.
\newblock {\em JHEP}, 04:080, 2017.

\bibitem{Buchmuller:2017vho}
Wilfried Buchmuller and Julian Schweizer.
\newblock {Flavor mixings in flux compactifications}.
\newblock {\em Phys. Rev. D}, 95(7):075024, 2017.

\bibitem{PhysRevD.97.075019}
Wilfried Buchmuller and Ketan~M. Patel.
\newblock Flavor physics without flavor symmetries.
\newblock {\em Phys. Rev. D}, 97:075019, Apr 2018.

\bibitem{Dixon:1985jw}
Lance~J. Dixon, Jeffrey~A. Harvey, C.~Vafa, and Edward Witten.
\newblock {Strings on Orbifolds}.
\newblock {\em Nucl. Phys. B}, 261:678--686, 1985.

\bibitem{Dixon:1986jc}
Lance~J. Dixon, Jeffrey~A. Harvey, C.~Vafa, and Edward Witten.
\newblock {Strings on Orbifolds. 2.}
\newblock {\em Nucl. Phys. B}, 274:285--314, 1986.

\bibitem{Atiyah:1963zz}
M.~F. Atiyah and I.~M. Singer.
\newblock {The index of elliptic operators on compact manifolds}.
\newblock {\em Bull. Am. Math. Soc.}, 69:422--433, 1969.

\bibitem{Witten:1984dg}
Edward Witten.
\newblock {Some Properties of O(32) Superstrings}.
\newblock {\em Phys. Lett. B}, 149:351--356, 1984.

\bibitem{Green:1987mn}
Michael~B. Green, J.~H. Schwarz, and Edward Witten.
\newblock {\em {SUPERSTRING THEORY. VOL. 2: LOOP AMPLITUDES, ANOMALIES AND PHENOMENOLOGY}}.
\newblock 7 1988.

\bibitem{Abe:2013bca}
Tomo-Hiro Abe, Yukihiro Fujimoto, Tatsuo Kobayashi, Takashi Miura, Kenji Nishiwaki, and Makoto Sakamoto.
\newblock {$Z_N$ twisted orbifold models with magnetic flux}.
\newblock {\em JHEP}, 01:065, 2014.

\bibitem{Abe:2014noa}
Tomo-hiro Abe, Yukihiro Fujimoto, Tatsuo Kobayashi, Takashi Miura, Kenji Nishiwaki, and Makoto Sakamoto.
\newblock {Operator analysis of physical states on magnetized $T^{2}/Z_{N}$ orbifolds}.
\newblock {\em Nucl. Phys. B}, 890:442--480, 2014.

\bibitem{Kobayashi:2017dyu}
Tatsuo Kobayashi and Satoshi Nagamoto.
\newblock {Zero-modes on orbifolds : magnetized orbifold models by modular transformation}.
\newblock {\em Phys. Rev. D}, 96(9):096011, 2017.

\bibitem{Kobayashi:2022tti}
Tatsuo Kobayashi, Hajime Otsuka, Makoto Sakamoto, Maki Takeuchi, Yoshiyuki Tatsuta, and Hikaru Uchida.
\newblock {Index theorem on magnetized blow-up manifold of T2/ZN}.
\newblock {\em Phys. Rev. D}, 107(7):075032, 2023.

\bibitem{Imai:2022bke}
Hiroki Imai, Makoto Sakamoto, Maki Takeuchi, and Yoshiyuki Tatsuta.
\newblock {Index and winding numbers on T2/ZN orbifolds with magnetic flux}.
\newblock {\em Nucl. Phys. B}, 990:116189, 2023.

\bibitem{Kikuchi:2022lfv}
Shota Kikuchi, Tatsuo Kobayashi, Kaito Nasu, and Hikaru Uchida.
\newblock {Classifications of magnetized T$^{4}$ and T$^{4}$/Z$_{2}$ orbifold models}.
\newblock {\em JHEP}, 08:256, 2022.

\bibitem{Kikuchi:2022psj}
Shota Kikuchi, Tatsuo Kobayashi, Kaito Nasu, Shohei Takada, and Hikaru Uchida.
\newblock {Number of zero-modes on magnetized T$^{4}$/Z$_{N}$ orbifolds analyzed by modular transformation}.
\newblock {\em JHEP}, 06:013, 2023.

\bibitem{Kikuchi:2023awm}
Shota Kikuchi, Tatsuo Kobayashi, Kaito Nasu, Shohei Takada, and Hikaru Uchida.
\newblock {Zero-modes in magnetized T6/ZN orbifold models through Sp(6,Z) modular symmetry}.
\newblock {\em Phys. Rev. D}, 108(3):036005, 2023.

\bibitem{Fujimoto:2016zjs}
Yukihiro Fujimoto, Tatsuo Kobayashi, Kenji Nishiwaki, Makoto Sakamoto, and Yoshiyuki Tatsuta.
\newblock {Comprehensive analysis of Yukawa hierarchies on $T^2/Z_N$ with magnetic fluxes}.
\newblock {\em Phys. Rev. D}, 94(3):035031, 2016.

\bibitem{Kikuchi:2021yog}
Shota Kikuchi, Tatsuo Kobayashi, Yuya Ogawa, and Hikaru Uchida.
\newblock {Yukawa textures in modular symmetric vacuum of magnetized orbifold models}.
\newblock {\em PTEP}, 2022(3):033B10, 2022.

\bibitem{Hoshiya:2022qvr}
Kouki Hoshiya, Shota Kikuchi, Tatsuo Kobayashi, and Hikaru Uchida.
\newblock {Quark and lepton flavor structure in magnetized orbifold models at residual modular symmetric points}.
\newblock {\em Phys. Rev. D}, 106(11):115003, 2022.

\bibitem{Sakamoto:2020vdy}
Makoto Sakamoto, Maki Takeuchi, and Yoshiyuki Tatsuta.
\newblock {Index theorem on $T^2/\mathbb{Z}_N$ orbifolds}.
\newblock {\em Phys. Rev. D}, 103(2):025009, 2021.

\bibitem{atiyah1967lefschetz}
Michael~Francis Atiyah and Raoul Bott.
\newblock A lefschetz fixed point formula for elliptic complexes: I.
\newblock {\em Annals of Mathematics}, pages 374--407, 1967.

\bibitem{Atiyah1968TheIndex}
M.~F. Atiyah and G.~B. Segal.
\newblock The index of elliptic operators: Ii.
\newblock {\em Annals of Mathematics}, 87(3):531--545, 1968.

\bibitem{berline1985computation}
Nicole Berline and Mich{\`e}le Vergne.
\newblock A computation of the equivariant index of the dirac operator.
\newblock {\em Bulletin de la Soci{\'e}t{\'e} math{\'e}matique de France}, 113:305--345, 1985.

\bibitem{bismut1985infinitesimal}
Jean-Michel Bismut.
\newblock The infinitesimal lefschetz formulas: a heat equation proof.
\newblock {\em Journal of functional analysis}, 62(3):435--457, 1985.

\bibitem{bismut1986localization}
J~M Bismut.
\newblock Localization formulas, superconnections, and the index theorem for families.
\newblock {\em Communications in mathematical physics}, 103:127--166, 1986.

\bibitem{mumford2007tata}
David Mumford and C~Musili.
\newblock {\em Tata lectures on theta. i (modern birkh{\"a}user classics)}.
\newblock Birkh{\"a}user Boston Incorporated, 2007.

\bibitem{Scherk:1978ta}
Joel Scherk and John~H. Schwarz.
\newblock {Spontaneous Breaking of Supersymmetry Through Dimensional Reduction}.
\newblock {\em Phys. Lett. B}, 82:60--64, 1979.

\bibitem{Atiyah:1964zzClifford}
M.~F. Atiyah, R.~Bott, and A.~Shapiro.
\newblock {Clifford modules}.
\newblock {\em Topology}, 3:S3--S38, 1964.

\bibitem{Bailin:1999nk}
D.~Bailin and A.~Love.
\newblock {Orbifold compactifications of string theory}.
\newblock {\em Phys. Rept.}, 315:285--408, 1999.

\bibitem{Takeuchi:2023egl}
Maki Takeuchi, Takanao Tsuyuki, and Hikaru Uchida.
\newblock {Three-generation solutions of equations of motion in heterotic supergravity}.
\newblock {\em Phys. Rev. D}, 107(9):095039, 2023.

\bibitem{Ishiguro:2020rot}
Keiya Ishiguro, Takafumi Kai, Satsuki Nishimura, Hajime Otsuka, and Maki Takeuchi.
\newblock {Upper bound on the Atiyah-Singer index from tadpole cancellation}.
\newblock {\em JHEP}, 24:200, 2020.

\end{thebibliography}

\end{document}